\documentclass[twocolumn,preprintnumbers,amsmath,amssymb,groupedaddress,superscriptaddress,pra]{revtex4-1}
\usepackage{amsmath,amsfonts,amssymb,graphicx,color,times,bbm}
\usepackage{graphicx}
\usepackage{dcolumn}
\usepackage{bm}
\usepackage{amssymb}
\usepackage{amsmath}
\usepackage[colorlinks=true,citecolor=blue,urlcolor=blue]{hyperref}

\usepackage{footmisc}
\renewcommand{\emph}{\textit}

\definecolor{mygray}{gray}{0.6}

\newcommand{\ket}[1]{| #1 \rangle}

\newcommand{\twotwo}{\ket{2,2}}
\newcommand{\twoone}{\ket{2,1}}

\newcommand{\uw}{$\mu$w}

\begin{document}
\title{Dual-species Bose-Einstein condensate of $^{41}$K and $^{87}$Rb in a hybrid trap}


\author{A. Burchianti}
\affiliation{Istituto Nazionale di Ottica, CNR-INO, 50019 Sesto Fiorentino, Italy}
\affiliation{LENS and Dipartimento di Fisica e Astronomia, Universit\`{a} di Firenze, 50019 Sesto Fiorentino, Italy}
\author{C. D'Errico}\email{derrico@lens.unifi.it}
\affiliation{Istituto Nazionale di Ottica, CNR-INO, 50019 Sesto Fiorentino, Italy}
\affiliation{LENS and Dipartimento di Fisica e Astronomia, Universit\`{a} di Firenze, 50019 Sesto Fiorentino, Italy}
\author{S. Rosi}
\affiliation{Istituto Nazionale di Ottica, CNR-INO, 50019 Sesto Fiorentino, Italy}
\affiliation{LENS and Dipartimento di Fisica e Astronomia, Universit\`{a} di Firenze, 50019 Sesto Fiorentino, Italy}
\author{A. Simoni}
\affiliation{Univ Rennes, CNRS, IPR (Institut de Physique de Rennes)-UMR 6251, F-35000 Rennes, France}
\author{M. Modugno}
\affiliation{Depto. de Fis\'ica Te\'orica e Hist. de la Ciencia, Universidad del Pais Vasco UPV/EHU, 48080 Bilbao, Spain}
\affiliation{IKERBASQUE, Basque Foundation for Science, 48013 Bilbao, Spain}
\author{C. Fort}
\affiliation{Istituto Nazionale di Ottica, CNR-INO, 50019 Sesto Fiorentino, Italy}
\affiliation{LENS and Dipartimento di Fisica e Astronomia, Universit\`{a} di Firenze, 50019 Sesto Fiorentino, Italy}
\author{F. Minardi}
\affiliation{Istituto Nazionale di Ottica, CNR-INO, 50019 Sesto Fiorentino, Italy}
\affiliation{LENS and Dipartimento di Fisica e Astronomia, Universit\`{a} di Firenze, 50019 Sesto Fiorentino, Italy}
\affiliation{Dipartimento di Fisica e Astronomia, Universit\`{a} di Bologna, 40127 Bologna, Italy}


\begin{abstract}


We report on the production of a $^{41}$K-$^{87}$Rb dual-species Bose-Einstein condensate in a hybrid trap, consisting of a magnetic quadrupole and an optical dipole potential. After loading both atomic species in the  trap, we cool down $^{87}$Rb first by magnetic and then by optical evaporation, while $^{41}$K is sympathetically cooled by elastic collisions with $^{87}$Rb. We eventually produce two-component condensates with more than $10^5$ atoms and tunable species population imbalance. We observe the immiscibility of the quantum mixture by measuring the density profile of each species after releasing them from the trap.
\end{abstract}
 
\maketitle

\section{INTRODUCTION}

Multi-component quantum gases are ideal platforms to study fundamental phenomena arising from the mutual interaction between different constituents. These effects occur in many physical systems ranging from superfluid helium mixtures to multi-component superconductors and neutron matter \cite{AndreevBashkin,Buckley,Babaev}. Since the first experimental observations of a Bose-Einstein condensate (BEC) in dilute gases \cite{Cornell95,Daviset95,Hulet95}, many efforts have been dedicated to the realization of degenerate atomic mixtures using different hyperfine states of single atomic species \cite{Cornell97,Cornell98,FortMinardi00}, different isotopes \cite{Truscott2001,Schreck2001,McNamara2006,Wieman2008,Tey2010,Lu2011,Sugawa2011,Grimm2013,Salomon2014} or different elements \cite{Modugno2002,Hadzibabic2002,Roati2002,Zimmermann2005,Dieckmann2008,Thalhammer2008,Lercher2011,Cornish2011,Park2012,Repp2013,Schreck2013,wacker_tunable_2015,Wang2016,Roy2017,Ospelkaus2018,Ferlaino2018}. Bose-Bose, Bose-Fermi and Fermi-Fermi mixtures are now produced in many laboratories worldwide and they are currently explored as benchmarks for addressing complex problems in many-body physics including collective \cite{FortMinardi00,Salomon2014} and topological excitations \cite{Cornell2004,Hoefer2011,Yao2016}, phase separation \cite{Wieman2008,Cornish2011,Proukakis2018}, magnetism \cite{kuklov_counterflow_2003, altman_phase_2003, Ueda2013}, polarons \cite{hu_bose_2016, jorgensen_observation_2016}, quantum droplets \cite{cabrera_quantum_2018,Semeghini}, spin superfluidity \cite{Ueda2013,Fava2018} and spin supercurrents \cite{Hadzibabic2013,Abad2014}. Ultracold quantum mixtures have also been exploited to produce ground-state polar molecules \cite{Jin2008,Cornish2014,Nagerl2014,Dulieu2016,JunYe2018}. Thus, the development of effective techniques to produce large and deeply degenerate two-component quantum gases deserves special attention.

In this paper, we present a simple and efficient route to prepare a $^{41}$K-$^{87}$Rb dual-species BEC in a hybrid trap. This specific Bose-Bose mixture is experimentally appealing because accessible heteronuclear Feshbach resonances in its ground state enable the control of the interspecies interactions \cite{Thalhammer2008}. 
In early experiments, degeneracy was reached by evaporative cooling of $^{87}$Rb with microwave (\uw) radiation, and sympathetic cooling of $^{41}$K \cite{Modugno2001,Modugno2002,Thalhammer2008}. Thanks to the large interspecies collision rate, two-component BECs have been produced in both magnetic \cite{Modugno2002} and optical potentials \cite{Thalhammer2008}. Generally, optical or hybrid traps are preferred to purely magnetic ones, due to their higher flexibility. Our strategy uses state-of-the-art cooling techniques and, at the same time, brings together the advantages of both magnetic quadrupoles and optical traps \cite{lin_rapid_2009}. This enables the production of large superfluid mixtures of $^{41}$K-$^{87}$Rb in a simple and reliable set-up, with a large optical access. We start by loading both atomic species, prepared in the $\left|F=2,m_{F}=2\right\rangle$ state, in a magnetic quadrupole. $^{87}$Rb is cooled by driving the \uw~transition to the ground hyperfine state, while $^{41}$K is cooled by thermal contact with $^{87}$Rb. The atoms are then loaded into a crossed optical dipole trap (ODT) through an intermediate cooling stage in a hybrid potential. The latter is given by the magnetic quadrupole plus a ``dimple'' beam, whose focus is  shifted from the zero of the quadrupole to minimize the Majorana spin-losses \cite{lin_rapid_2009}. The final step is a pure optical evaporation in the ODT, created by crossing the ``dimple'' with an auxiliary beam. Within an experimental cycle of less than 20~s, we produce stable dual-species BECs with more than $10^5$ atoms and tunable species population imbalance. This result represents a convenient starting point for future studies on mass-imbalanced superfluid mixtures with tunable interactions which are expected to exhibit exotic phenomena such as the formation of unusual vortex structures \cite{Barnett2008,Kuopanportti2012,Kuopanportti2015}, self-bound states \cite{Petrov2015} and non-dissipative drag effects \cite{AndreevBashkin,Shevchenko2005,Nespolo2017,Babaev2018,Giorgini2018}.

The article is organized as follows. In Section \ref{sec:Experiment}, we briefly describe our setup. In Section \ref{sec:hybrid}, we report the cooling of the atomic mixture in the hybrid trap. In Section \ref{sec:klosses}, we experimentally investigate the lifetime of both species during the \uw~evaporation. In Section \ref{sec:dualBEC}, we detail the creation of the dual-species condensate, we observe its immiscibility and compare our experimental results with the prediction of the mean field theory. Finally, in Section \ref{sec:conclusions} we draw the conclusion.

\section{EXPERIMENT}
\label{sec:Experiment}

\begin{figure}[t]
\includegraphics[width = .5\textwidth]{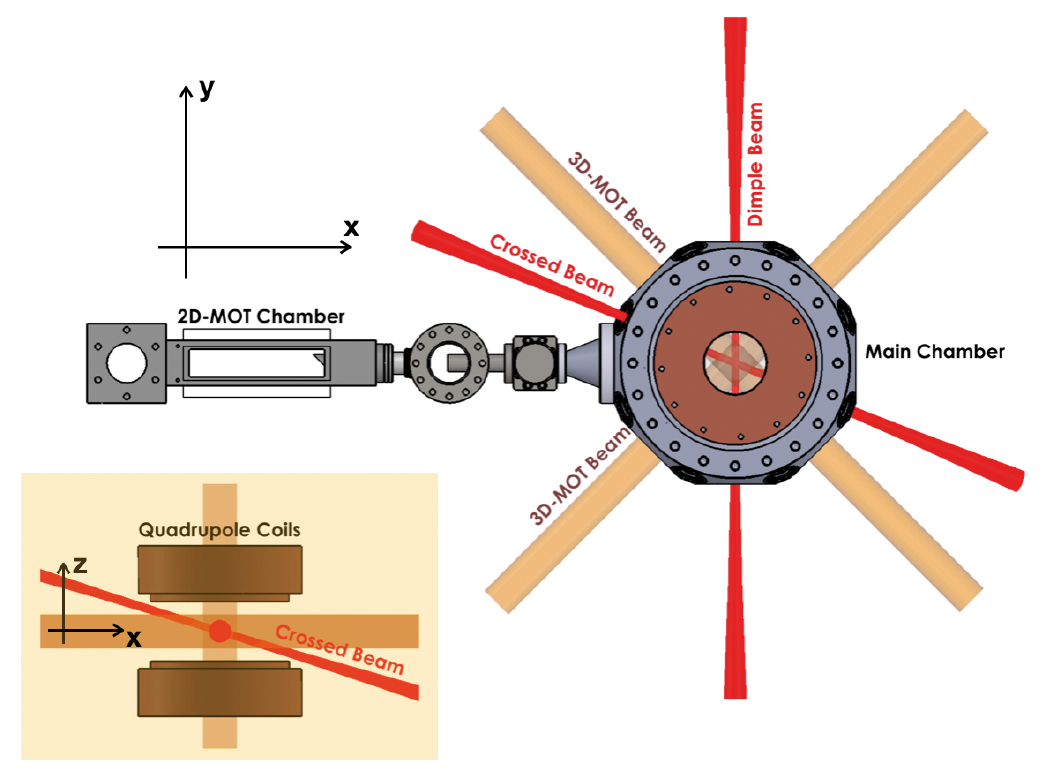}
\caption{\label{fig:apparatus} Top view of the experimental apparatus, showing the 2D-MOT and the ``science'' chamber. The 3D-MOT (ODT) beams are also depicted in orange (red). Inset: schematic front view of the quadrupole coils and the beams. The ``dimple'' beam is directed along the $\hat{y}$ axis. The ``crossed'' beam forms an angle of 67.5$^\circ$ with the ``dimple'' beam in the $xy$ plane and is inclined at an angle of 16$^\circ$ with respect to that plane.}
\end{figure}

\begin{figure}[t]
\includegraphics[width = .5\textwidth]{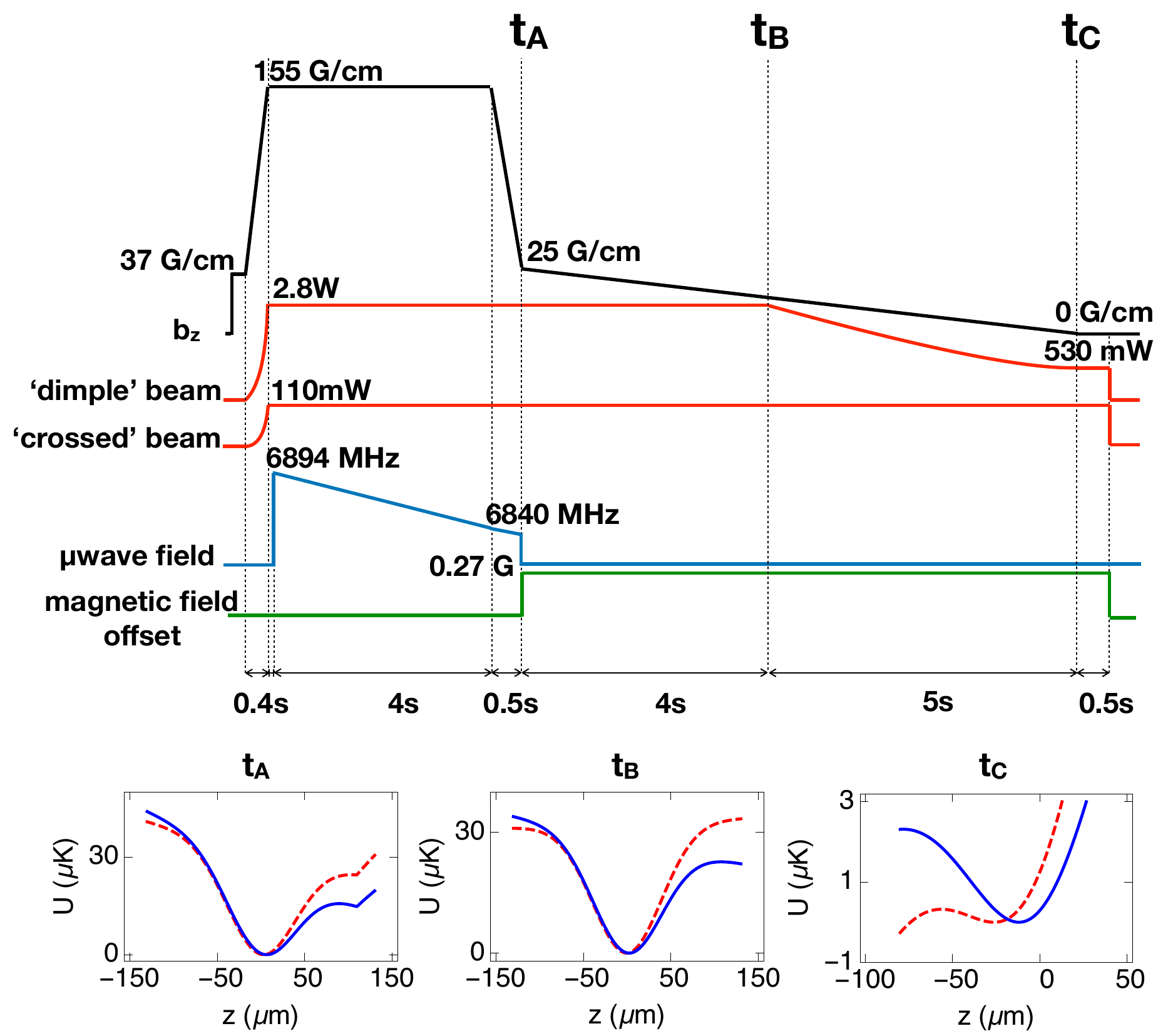}
\caption{\label{fig:sequence} Upper panel: temporal sequence of the evaporative/sympathetic cooling leading to dual-species BEC. Lower panel: hybrid trap potential, along the $\hat{z}$ axis, for $^{41}$K (blue solid line) and $^{87}$Rb (red dashed line), corresponding to different times. The $z=0$ position corresponds to the initial quadrupole center, on which the dipole trap is aligned. At $\rm{t_{A}}$ the position of the zero magnetic field is vertically shifted to avoid Majorana spin-flips, at $\rm{t_{B}}$ the optical evaporation starts and at the end of the evaporation, $\rm{t_{C}}$, only a weak optical trap remains, with a vertical gravitational sag of $\sim 14~\mu$m between the two species.}
\end{figure}

The core of the experimental setup is schematically shown in Fig.~\ref{fig:apparatus}. It consists of two main parts: the two-dimensional magneto-optical trap (2D-MOT) chamber, where we produce a cold atomic beam of both $^{41}$K and $^{87}$Rb, and the three-dimensional magneto-optical trap (3D-MOT) ``science'' chamber, where we produce the dual-species condensate. These two parts are connected through a differential pumping section providing a low conductance between them. The background vapor pressure in the 2D-MOT chamber is $\sim10^{-8}$ mbar, while in the 3D-MOT chamber is $\sim10^{-11}$ mbar. The 2D-MOT is loaded by a thermal vapor of K and Rb released in natural abundance by two metallic reservoirs (see Fig.~\ref{fig:apparatus}). 
Once transversely cooled by the 2D-MOT, the atoms are pushed towards the ``science'' chamber by resonant push beams. Here, they are captured by a standard 3D-MOT, consisting of six independent beams (shown in orange in Fig.~\ref{fig:apparatus}). We approximately load $2 \times10^7$  ($5 \times10^9$ ) atoms/s of $^{41}$K  ($^{87}$Rb ). To control the number of atoms in the MOT, we stabilize the MOT fluorescence signal by actively adjusting the push beams intensity level. After a compressed-MOT and a molasses phases, both atomic species are pumped in the $\ket{2,2}$ low-field seeking state and are magnetically captured in a quadrupole magnetic field, generated by the same coils used for the 3D-MOT. These coils are placed along the vertical $\hat{z}$ axis within reentrant viewports, above and below the science chamber. The quadrupole axial gradient is raised to the value $b_z=37$~G/cm (see Fig.~\ref{fig:sequence}), sufficiently high to hold the heavier $^{87}$Rb atoms against gravity, and then is ramped to its maximum value $b_z=155$~G/cm, together with two off-resonant laser beams at a wavelength of 1064~nm (shown in red in Fig~\ref{fig:apparatus}). The ``dimple'' beam, directed along the $\hat{y}$ axis, has a power of 2.8~W and waists $w_x$ and $w_z$ of 115~$\mu$m and 75~$\mu$m, respectively. The weaker ``crossed'' beam, with a waist of 70~$\mu$m, has a power of 110~mW. The latter beam crosses the ``dimple'' beam at an angle of 67.5$^\circ$ in the horizontal $xy$ plane and is inclined at an angle of 16$^\circ$ with respect to the same plane (see Fig.~\ref{fig:apparatus}). These two red-detuned focused beams intersect at the center of the quadrupole magnetic field.

\section{EVAPORATIVE AND SYMPATHETIC COOLING}
\label{sec:hybrid}
We load in the compressed quadrupole about $3\times 10^7$ atoms of $^{41}$K at 1~mK, and $4\times 10^9$ atoms of $^{87}$Rb at 300~$\mu$K. To further decrease the temperature, $^{87}$Rb is cooled first by magnetic and then by optical evaporation (see Fig.~\ref{fig:sequence}), while $^{41}$K is sympathetically cooled via elastic collisions with $^{87}$Rb. The magnetic evaporation is performed by means of a selective \uw~radiation around 6.8~GHz driving the $^{87}$Rb hyperfine transition $\ket{2,2}\rightarrow\ket{1,1}$; in 4.5~s the energy cut is linearly ramped from 1.9 to 0.17~mK.
At this point, the temperature is approximately 30 $\mu$K and the Majorana losses become significant. Thus the  \uw~radiation is switched off and the magnetic field gradient is decompressed down to $b_z=25$~G/cm in 0.5~s, thereby adiabatically cooling the gas below 10 $\mu$K. Then, we add a magnetic bias field to vertically shift
  the zero of the quadrupole from the center of the dipole trap to
  $\Delta z_Q=0.1$~mm above it ($\rm{t_{A}}$ in Fig.~\ref{fig:sequence}). The evaporation
  is continued by lowering $b_z$ to zero in 9~s, which loads the atoms into the
  purely optical trap and increases $\Delta z_Q$  inversely
  proportional to $b_z$. Since the depth of our optical trap is only 30 $\mu K$, extinguishing the magnetic confinement causes a drop of the atom numbers and the temperature. Finally, we reduce the intensity of the ``dimple'' beam from 2.8~W to 0.53~W in 5~s, while the ``crossed'' beam remains at full power, as shown in
  Fig.~\ref{fig:sequence}.



\begin{figure}[t]
\includegraphics[width = .45\textwidth]{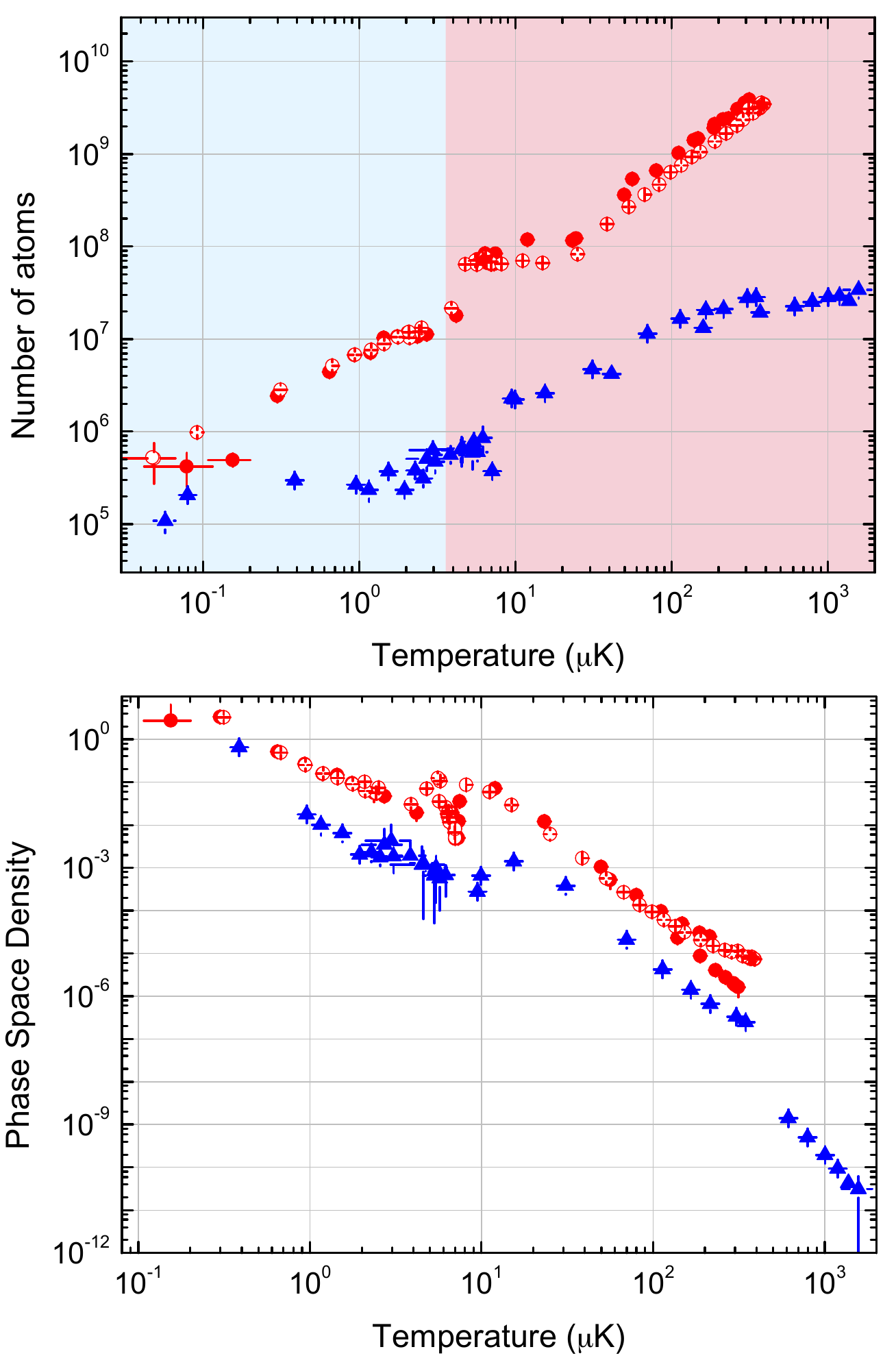}
\caption{\label{fig:evaporationPlots} Cooling trajectory: number of atoms (upper panel) and phase-space density (lower panel) of $^{41}$K (blue filled triangles) and $^{87}$Rb (red filled circles) versus the temperature $T$, during evaporation. For comparison we also report the data obtained for the evaporation of a $^{87}$Rb sample in the absence of $^{41}$K (red empty circles). The light blue (red) area corresponds to the optical (magnetic) evaporation.}
\end{figure}

In Fig.~\ref{fig:evaporationPlots} we show the number of atoms (upper panel) and the phase space density (lower panel) of both $^{41}$K (blue filled triangles) and $^{87}$Rb (red filled circles) as a function of the temperature $T$, measured during the cooling ramp. The red region corresponds to the magnetic evaporation (from the switching-on of the \uw\ power to $\rm{t_{B}}$ in Fig.~\ref{fig:sequence}), while the light blue region corresponds to the optical evaporation (from $\rm{t_{B}}$ to $\rm{t_{C}}$ in Fig.~\ref{fig:sequence}). In both regions, the $^{87}$Rb\ atom number is reduced by about two orders of magnitude. For comparison, we also report the atom number and the phase space density measured by loading only $^{87}$Rb\ into the hybrid trap (red empty circles). No appreciable differences are observed in the $^{87}$Rb\ evaporation trajectory, with or without $^{41}$K, at least above 300~nK, proving that the  thermal load, due to $^{41}$K, is too small to affect the evaporation efficiency of $^{87}$Rb. The $^{41}$K\ atom number is, in fact,  from two to one order of magnitude lower than $^{87}$Rb\, except at the end of evaporation when they become comparable.

Even if the \uw\ radiation selectively removes only $^{87}$Rb\ atoms, we observe significant losses of the $^{41}$K\ population, which decreases by more than one order of magnitude during the magnetic evaporation. In the next section, we will focus on this specific issue, inferring that the $^{41}$K\ lifetime in the compressed quadrupole is limited by a residual fraction of $^{87}$Rb\ atoms in the $\twoone$ state, as already noted in Refs.~\cite{Modugno2002,PhysRevLett.106.240801,wacker_tunable_2015}. Here, we just point out that, in previous experiments, such losses were severe enough to prevent dual-species condensation unless the $^{87}$Rb $\twoone$ atoms were continuously removed from the trap. However, standard ``cleaning'' strategies, based on the addition of a second \uw\ radiation to eliminate the undesired $^{87}$Rb population \cite{Modugno2002,Zimmermann2005,de_sarlo_collisional_2007,campbell_efficient_2010,wacker_tunable_2015}, are not viable in our case, due to the zero minimum of the quadrupole field. Despite this drawback, we find that the efficiency of the hybrid trap still assures the production of large dual-species condensates.


When the optical evaporation starts, we typically have $5 \times 10^5$ atoms of $^{41}$K and $2 \times 10^7$ atoms of $^{87}$Rb at a temperature of a few $\mu$K. Due to the larger mass, the trap is shallower for $^{87}$Rb than for $^{41}$K, thus, the optical evaporation mainly removes $^{87}$Rb atoms and sympathetically cools $^{41}$K, as confirmed by the almost horizontal slope of the $^{41}$K cooling trajectory (see Fig. \ref{fig:evaporationPlots} (a)). At this stage of evaporation, the efficiency of sympathetic cooling is sustained by the large interspecies scattering length $a_{12}=163 a_{0}$ \cite{Catani2008}. At the end, when the temperature is below $\sim$ 300~nK, the trap becomes shallow enough to evaporate $^{41}$K too. We observe that, in order to produce almost pure dual-species condensates, the auxiliary ``crossed'' beam is necessary to maintain a sufficient spatial overlap between the two atomic clouds. Otherwise, the $^{41}$K sample is only partially condensed. Differently, in the absence of $^{41}$K, we can produce pure $^{87}$Rb condensates using only the ``dimple'' beam, with the confinement in the $y$ direction provided by a non-zero quadrupole field of $9$ G/cm. However, to simultaneously transfer both species from the quadrupole to the dipole trap, the above quadrupole field gradient is too large, since it generates a secondary potential minimum trapping K atoms around the zero of the magnetic field. The parameters of the dipole trap are not critical: the power of the “crossed” (“dimple”) can be varied from 50 to 200 mW (600 to 300 mW), thus providing average trap frequencies in the range 90 - 100 Hz (50 - 60 Hz) for $^{41}$K ($^{87}$Rb).  In particular we need frequencies larger than about 40 (50) Hz in each direction for $^{41}$K ($^{87}$Rb), to avoid the spatial separation of the two species.

\section{INELASTIC COLLISIONS IN THE MAGNETIC TRAP}
\label{sec:klosses}

\begin{figure}[t]
\includegraphics[width = 0.5\textwidth]{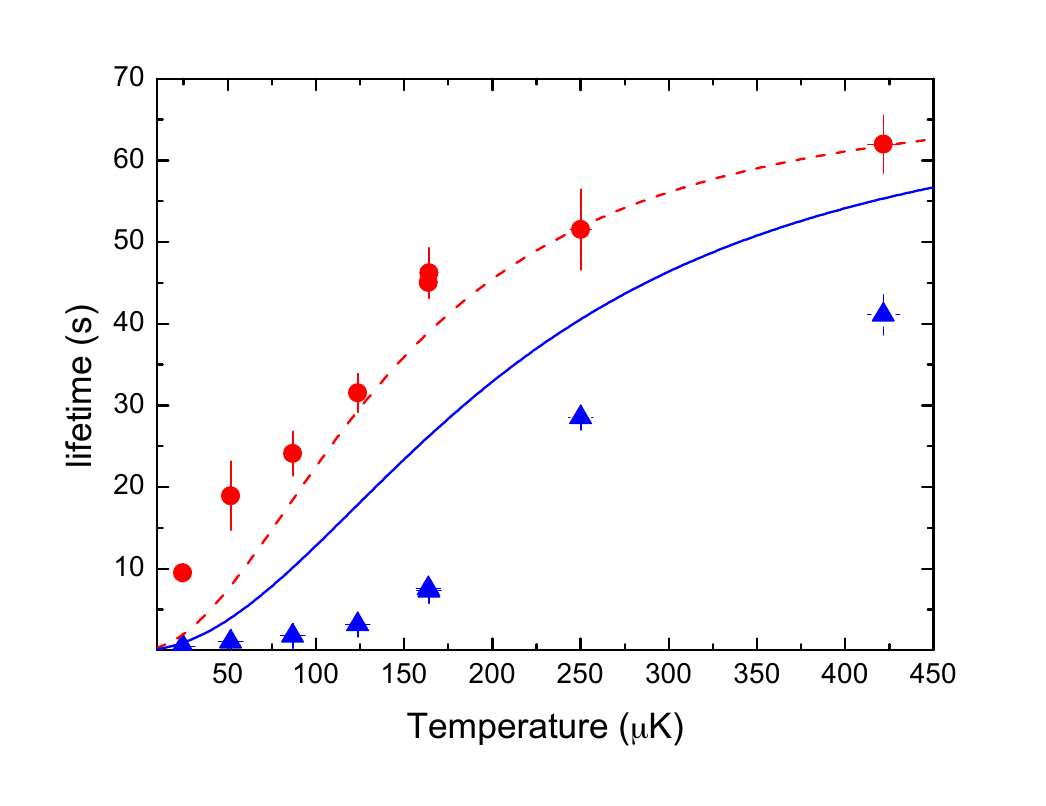}
\caption{\label{fig:lifetimes} Lifetimes of $^{41}$K (blue filled triangles) and of $^{87}$Rb (red filled circles) measured in absence of \uw\ radiation for different temperatures reached during the magnetic evaporation in the hybrid trap. Lines correspond to the function $\tau_{\mathrm{fit}}$, reported in the text, for $^{41}$K (blue solid) and $^{87}$Rb (red dashed), with the fit parameters $\Gamma_\mathrm{bg}=0.016(1)$ s$^{-1}$ and $\chi = 0.21\pm0.05$ obtained from$^{87}$Rb data (uncertainties equal to 1$\sigma$). The deviation between the fit function and the $^{41}$K data  suggests that background collisions and Majorana spin-flips are not the only mechanisms responsible for the observed losses of $^{41}$K atoms (see text). }
\end{figure}

\begin{figure}[t]
\includegraphics[width = 0.5\textwidth]{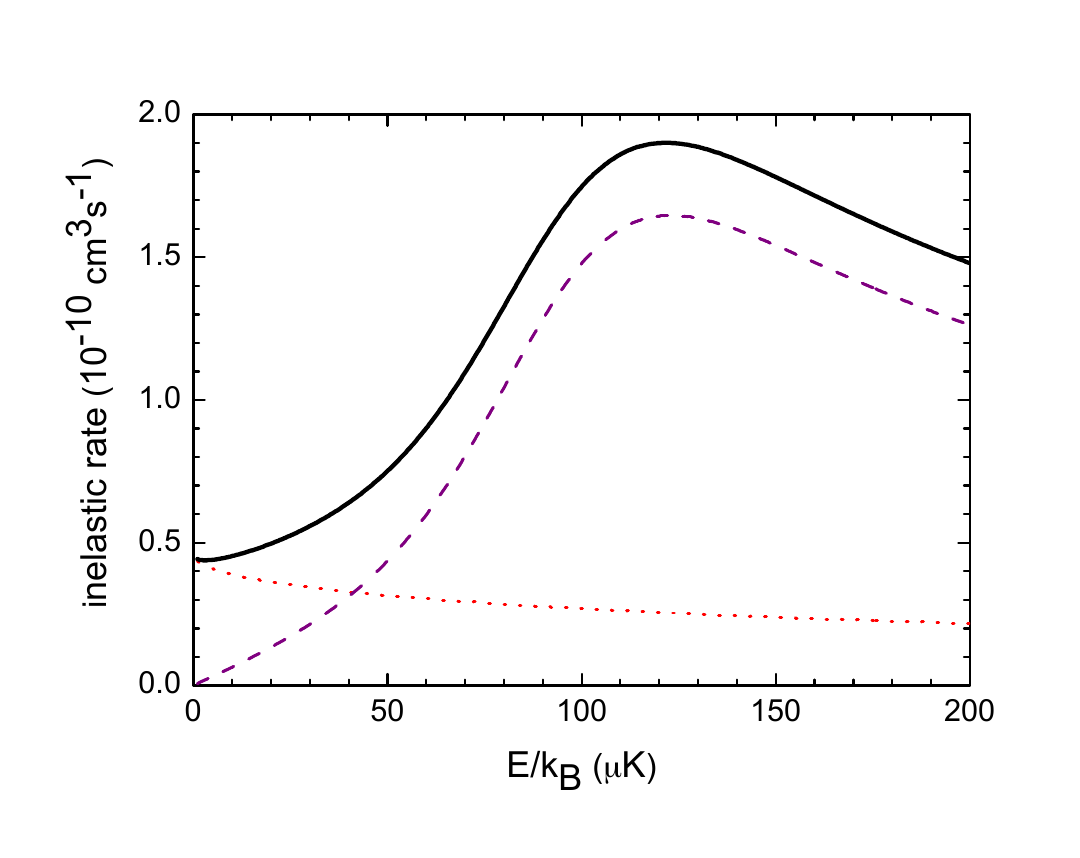}
\caption{\label{fig:collisionrate} Calculated inelastic collision rate between $^{41}$K $\twotwo$ and $^{87}$Rb $\twoone$ atoms. Red dotted line: $s$-wave contribution. Purple dashed line: $p$-wave contribution, including a factor of three arising from the sum over orbital angular momentum projections. Black solid line: total. The calculation is performed at a magnetic field of 0.5~G (for magnetic fields in the range of a few gauss no substantial changes are observed).}
\end{figure}

In this section, we investigate the causes underlying the $^{41}$K losses observed during the magnetic evaporation. To this end, we measure the lifetime of both atomic species in the compressed quadrupole. We halt the evaporation at intermediate times, by switching off the \uw\ power, and hereafter we measure the number of atoms decaying in time. We fit all decays with simple exponential functions and we report in Fig.~\ref{fig:lifetimes} the lifetimes $\tau_\mathrm{K}$ and $\tau_\mathrm{Rb}$ as a function of the temperature $T$, which decreases during the evaporation. First, we explore the possibility that the observed lifetimes are due to nonadiabatic transitions towards magnetically untrapped states, namely Majorana spin-flips. In a quadrupole trap, the Majorana loss rate is given by $\Gamma_\mathrm{m}=\chi (\hbar/M) (\mu b_z/ k_B T)^2$, with $\hbar$ the reduced Planck constant, $M$ the atomic mass, $\mu$ the atomic magnetic moment and $k_B$ the Boltzmann constant \cite{Petrich1995}. The dimensionless factor $\chi$ can be directly evaluated from the data and it has been found to be 0.16 for Rb \cite{Dubessy2012}, and 0.14 for Na \cite{Heo2011}. Thus, we compare the measured lifetimes with the expected trend. We first fit $\tau_\mathrm{Rb}$  with the function $\tau_{\mathrm{fit}} = (\Gamma_\mathrm{bg}+\Gamma_\mathrm{m})^{-1} $, taking also into account the one-body loss rate $\Gamma_\mathrm{bg}$ due to collisions with the background gas. Here, $\Gamma_\mathrm{bg}$ and $\chi$ are fitting parameters. We find that the values of $\tau_\mathrm{Rb}$ are qualitatively reproduced by $\tau_{\mathrm{fit}}$, with $\chi = 0.21(5)$, consistent with Ref. \cite{Dubessy2012}, and $\Gamma_\mathrm{bg}=0.016(1)$ s$^{-1}$. Now, assuming the same values extracted from the $^{87}$Rb data, the Majorana loss rate for $^{41}$K atoms should increase by a factor $M_\mathrm{Rb} /M_\mathrm{K}$ for each value of $T$. Nevertheless, as shown in Fig.~\ref{fig:lifetimes}, this gives an overestimation of $\tau_\mathrm{K}$. It follows that background collisions and Majorana spin-flips are not the only mechanisms involved.

As already mentioned in the previous section, an additional source of the observed $^{41}$K losses is the presence of a residual fraction of $^{87}$Rb $\twoone$ atoms, which can drive, via fast spin-exchange collisions, $^{41}$K $\twotwo$ atoms into the magnetically untrapped $\ket{1,1}$ state. We calculate this inelastic collision rate, using the predictive model developed in Ref.~\cite{simoni2008}. In the range of temperatures here explored, we find that it varies from $5 \times 10^{-11}$ to $1.9 \times 10^{-10}$ cm$^3$/s, as shown in Fig.~\ref{fig:collisionrate}. We confirm the presence of $^{87}$Rb $\twoone$ atoms by measuring the $^{87}$Rb spin composition via Stern-Gerlach separation induced by a magnetic field gradient, during time-of-flight (TOF) expansion. This is feasible only at temperatures below 50~$\mu$K, which are reached at the end of the \uw\ evaporation. We find that approximately 10\% of the $^{87}$Rb\ atoms are in the $\twoone$ state. At this cooling stage, using the calculated collision rate and the measured $^{41}$K $\twotwo$ and $^{87}$Rb $\twoone$ atom numbers, we estimate the $1/e$-decay time of $^{41}$K atoms to be approximately half a second, 
in agreement with the experiment.

Although the initial magnetic field gradient is unable to sustain $^{87}$Rb\ atoms in $\twoone$, this state can be continuously populated, during the magnetic evaporation, by several mechanisms such as: {\it (i)} Majorana spin-flips $\twotwo \rightarrow \twoone$; {\it (ii)} dipolar collisions $\twotwo +\twotwo \rightarrow \twoone + \twotwo$ \footnote{We develop a $^{87}$Rb collision model based on the accurate potential parameters determined in \cite{vanKepen2002}. In addition to the well-known dipolar coupling, our model also includes the second-order spin-orbit interaction, first discussed in the context of ultracold gases in \cite{Krauss1996}. The atom-loss rate is found to be sensitive to the variation of collision energy and magnetic field, and in the range of our experimental parameters is predicted to vary between $1 \times 10^{-15}$ and $2 \times 10^{-14}$ cm$^3$/s. } ; {\it (iii)} \uw\ photons, absorbed by the evaporated $\ket{1,1}$ atoms while leaving the trap and reaching regions of higher magnetic field \cite{Meschede2007}. Actually, we cannot single out one dominant effect as they appear to be of the same order of magnitude: each of them can produce enough $^{87}$Rb $\twoone$ atoms to cause severe losses in $^{41}$K.  A deeper understanding of the processes causing the transfer of Rb atoms in the $\twoone$ state will deserve further investigations which are beyond the scope of this work.

\section{DUAL-SPECIES BOSE-EINSTEIN CONDENSATE}
\label{sec:dualBEC}

\begin{figure}[t]
\includegraphics[width = 0.45\textwidth]{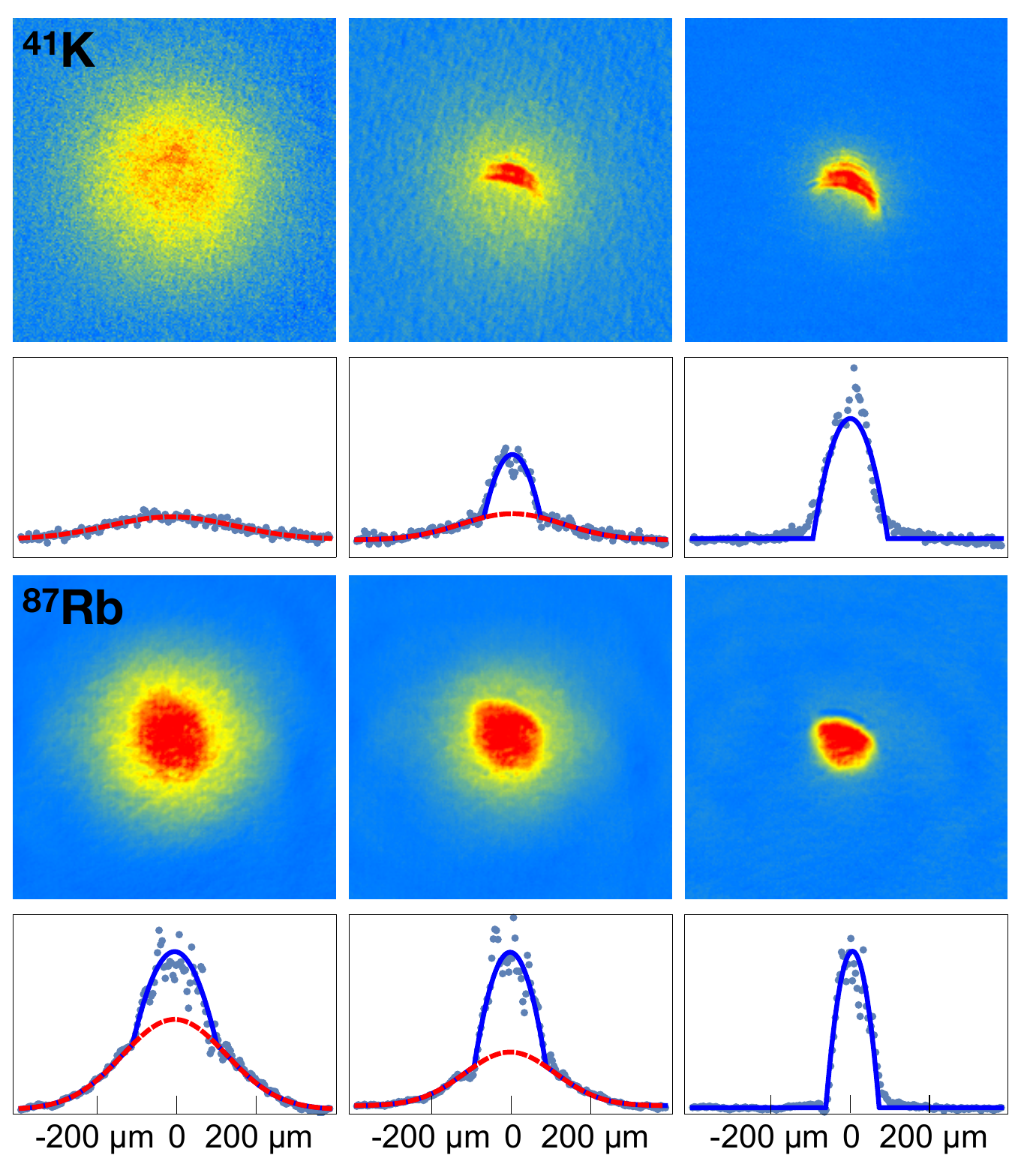}
\caption{\label{fig:transizione} Absorption images (776 $\mu$m x 776 $\mu$m) and line densities of the atomic clouds near the dual-species BEC phase transition. Images are taken after 18~ms of TOF for $^{41}$K (top) and 26~ms for $^{87}$Rb (bottom). Blue solid lines are fitting results with a two-component function: a Thomas-Fermi profile plus a Gaussian function. Thermal components (red dashed lines) and temperature decrease as the ``dimple'' beam power decreases: 0.92~W ($N^0_\mathrm{K}/N_\mathrm{K} \simeq$ 0, $N^0_\mathrm{Rb}/N_\mathrm{Rb} \simeq$ 0.1, T $\sim$ 300~nK), 0.73~W ($N^0_\mathrm{K}/N_\mathrm{K} \sim$ 0.09, $N^0_\mathrm{Rb}/N_\mathrm{Rb} \simeq$ 0.24, T $\sim$ 250~nK) and 0.53~W ($N^0_\mathrm{K} \simeq 6 \times 10^4$, $N^0_\mathrm{Rb} \simeq 4 \times 10^5$) (from left to right).}
\end{figure}

Once that the atomic mixture has been transferred into the ODT, the degenerate regime is reached by lowering the trap depth. In Fig.~\ref{fig:transizione} we show the density profiles of $^{41}$K and $^{87}$Rb at different stages of the optical evaporation across the BEC phase transition. The images of both atomic clouds are taken by absorption imaging in the $xz$ plane, after switching off the two trapping beams. We observe that the condensation is reached first for $^{87}$Rb and then for $^{41}$K. In fact, even if the estimated $^{41}$K  trap frequencies are about a factor 1.4 larger than the $^{87}$Rb ones, the $^{87}$Rb atom number exceeds $^{41}$K by more than one order of magnitude at the condensation threshold. At the end of evaporation, when no thermal component is discernible anymore, we have $N^0_\mathrm{K} \simeq 6 \times 10^4$ and $N^0_\mathrm{Rb} \simeq 4 \times 10^5$ atoms  \footnote{The atom number $N$ of both $^{41}$K and $^{87}$Rb  have been calibrated using the saturation absorption imaging technique, described in \cite{Reinaudi:07}. We estimate an  uncertainty  in $N$  of 35\%  for both atomic species.}.  

\begin{figure}[t]
\includegraphics[width=0.48\textwidth]{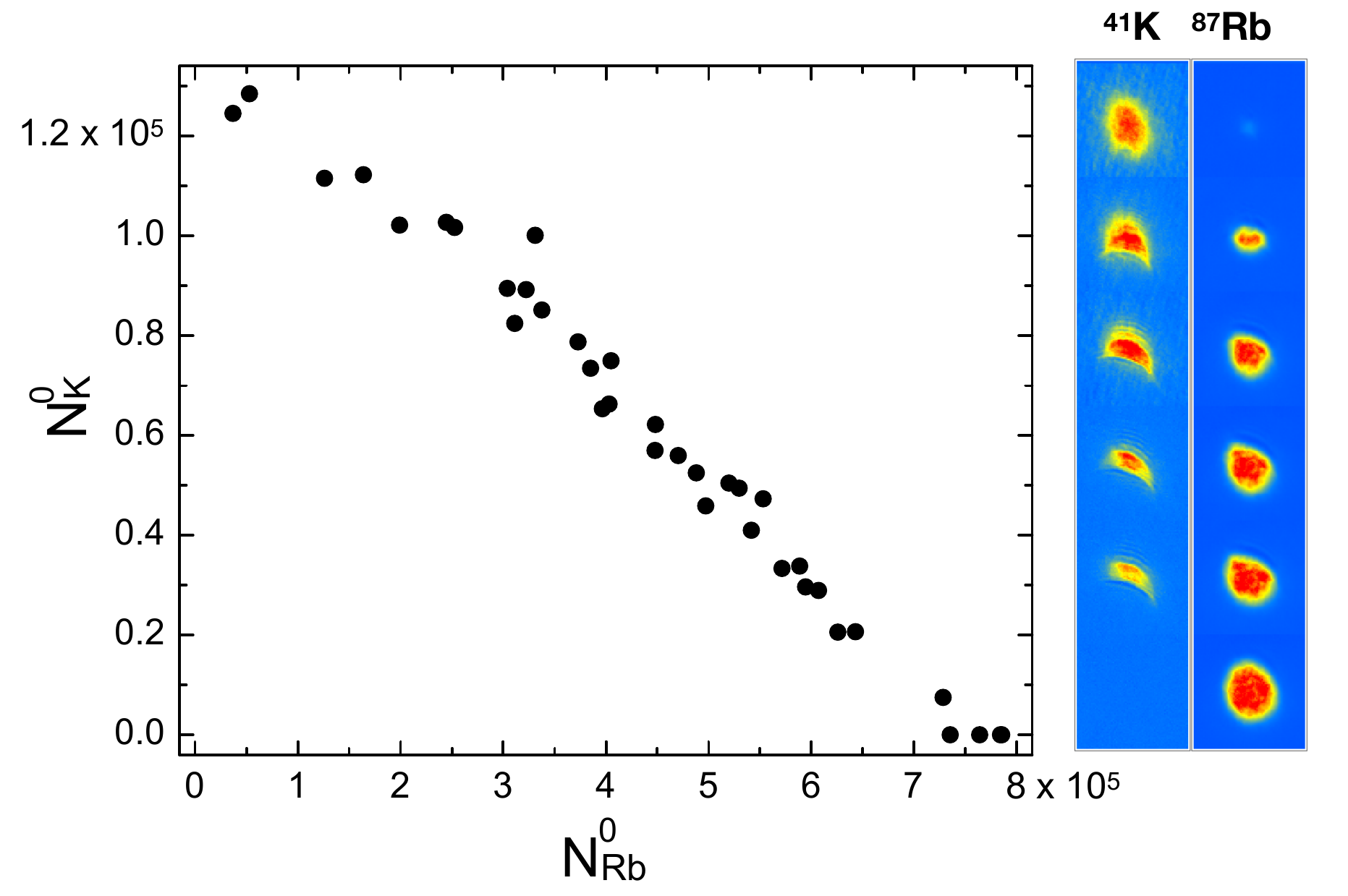}
\caption{\label{fig:NK_vs_NRb}  Number of $^{41}$K condensed atoms, $N^0_\mathrm{K}$, as a function of the number of $^{87}$Rb condensed atoms, $N^0_\mathrm{Rb}$. Their ratio is tuned by adjusting the number of $^{41}$K and $^{87}$Rb atoms initially loaded in the magnetic quadrupole. Inset: Absorption images (310 $\mu$m x 310 $\mu$m) of $^{41}$K (left) and $^{87}$Rb (right) BECs for different values of the species population imbalance. Images are taken after 18~ms of TOF for $^{41}$K and 21~ms for $^{87}$Rb. }
\end{figure}

\begin{figure}[t]
\includegraphics[width = .39\textwidth]{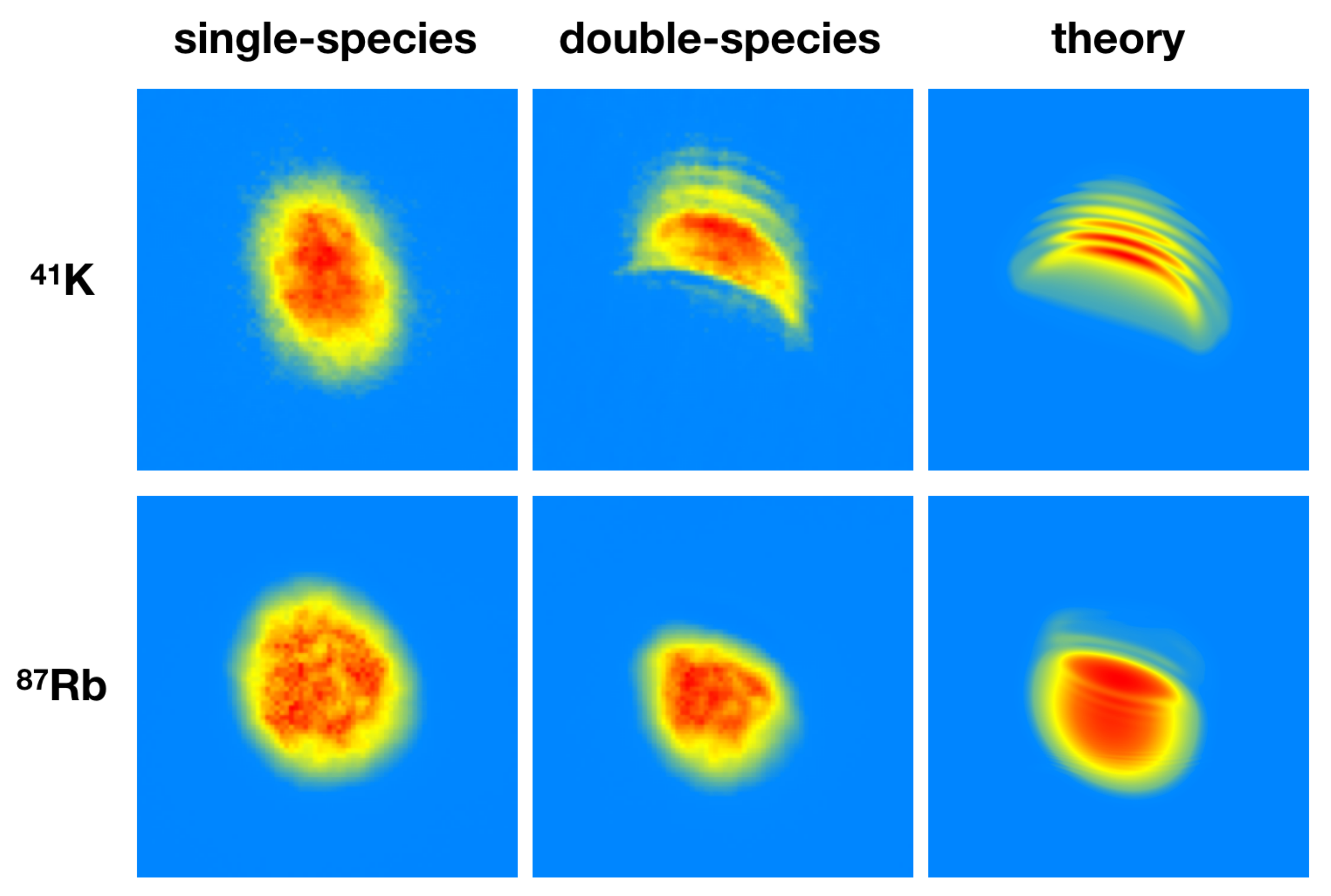}
\caption{\label{fig:with_without_TH} Absorption images (310 $\mu$m x 310 $\mu$m) of single-species (left column) and dual-species condensates (middle column). Images are taken after 18~ms TOF for $^{41}$K (top) and 21~ms for $^{87}$Rb (bottom). $N^0_\mathrm{K} \simeq 8 \times 10^4$ ($\simeq 7 \times 10^4$) in the case of single(double)-species BEC. $N^0_\mathrm{Rb} \simeq 9 \times 10^5$ ($\simeq 5 \times 10^5$) in the case of single(double)-species BEC. Right column: simulated density profiles of the two-interacting BECs. The simulation (see text) is performed using the same experimental parameters corresponding to the central image. The center of all images corresponds to the position of the expanding single-species BECs. The tilt between the two phase-separated condensates, observed both in the experiment and in the simulation, is due to a few microns shift of their in-trap centers along the $\hat{x}$ axis. Such effect is caused by the oblique direction of the "crossed" beam.}
\end{figure}

As shown in Fig.~\ref{fig:NK_vs_NRb}, the population imbalance of the double-species BEC can be tuned, by changing the ratio of $^{41}$K and $^{87}$Rb atoms initially loaded in the magnetic quadrupole. As a natural consequence of sympathetic cooling, we observe that the number of condensed atoms $N^0_\mathrm{K}$ and $N^0_\mathrm{Rb}$ are anticorrelated. Once $N_\mathrm{K}$ increases, the larger thermal load causes larger evaporation losses on the coolant species, i.e. $N^0_\mathrm{Rb}$ decreases. On the other hand, once $N_\mathrm{Rb}$ increases, the $^{41}$K losses due to spin-exchange collisions with residual $^{87}$Rb $\twoone$ atoms rise, i.e. $N^0_\mathrm{K}$ decreases. Using the same cooling sequence, we also produce single-species $^{87}$Rb BECs with $N^0_\mathrm{Rb} \simeq 8 \times 10^5$ atoms. In order to condense $^{41}$K, without any residual component of $^{87}$Rb, instead, we further decrease the trap depth, by reducing the power of both ODT beams. In this way, after evaporating all $^{87}$Rb atoms, direct optical evaporation of  $^{41}$K proceeds, ending with the production of single-species $^{41}$K BECs with $N^0_\mathrm{K} \simeq 1.5 \times 10^5$ atoms (not shown in Fig.~\ref{fig:NK_vs_NRb}).
 
In Fig.~\ref{fig:with_without_TH} we directly compare the TOF absorption images of single and dual-species condensates. We find that, in the latter case, the density distribution of each species is affected by the presence of the other. In particular, the lower part of $^{41}$K and the upper part of $^{87}$Rb repeals each other, and their vertical separation increases by a few tens of $\mu$m  with respect to the positions of the single-species BECs. This behavior indicates a strong repulsive interspecies interaction. Within the Thomas-Fermi approximation, the ground-state of two interacting BECs can be described in terms of the coupling constants \cite{MicheleModugno2002}: $g_{ij}=\frac{2\pi \hbar^{2} a_{ij}}{M_{ij}}$, with  $M_{ij}=\frac{M_{i}M_{j}}{\left(M_{i}+M_{j}\right)}$ ($i,j=1,2$). Here, $a_{ii}$ and $M_{ii}$ are the intraspecies scattering length and the atomic mass of the $i$ species, while $a_{ij}$ and $M_{ij}$ are the interspecies scattering length and the reduced mass. For our quantum mixture, we have: $a_{11}=65 a_{0}$,  $a_{22}=99 a_{0}$ and $a_{12}=163 a_{0}$ \cite{Catani2008}, where $a_{0}$ is the Bohr radius and $^{41}$K ($^{87}$Rb) is labeled as species 1 (2). Thus, since the relation $g_{12} > \sqrt{g_{11}g_{22}}$ is fulfilled, we expect that the two components are phase-separated. However, due to the gravitational sag, the in-trap Thomas-Fermi density distributions hardly overlap. This suggests, as already observed in \cite{wacker_tunable_2015,Proukakis2018}, that the repulsion effect arises during the TOF expansion, once the two clouds start to overlap. We verify this hypothesis by numerically solving a system of two coupled Gross-Pitaevskii equations, which describe the two-component BEC expanding from the trap \footnote{A set of two three-dimensional Gross-Pitaevskii (GP) equations is considered \cite{Pethick}. The ground state is found by using a standard imaginary time evolution \cite{DalfovoRev1999}. The time-dependent GP equations are solved by means of a split-step method that makes use of fast Fourier transforms \cite{Jackson1998}}.  We perform the simulation using the same experimental parameters corresponding to the dual-species BEC in Fig.~\ref{fig:with_without_TH}. We find good agreement between experiment and theory: not only the density distribution of the two BECs but also the shift of their centers of mass are well reproduced, confirming our expectation. In the future, we will explore the miscible phase diagram in the lowest hyperfine states where convenient Feshbach resonances allow for fine tuning of $a_{12}$. 

\section{CONCLUSIONS}

\label{sec:conclusions}

We have demonstrated fast and efficient production of $^{41}$K-$^{87}$Rb dual-species condensates in a hybrid trap. The method is based on  evaporative/sympathetic cooling in  magnetic quadrupole and optical potentials. The magnetic quadrupole allows the loading of large atomic samples, while the optical trap provides fast evaporation and thermalization of the atomic mixture, approaching the degenerate regime.
Even though we observe severe losses of $^{41}$K atoms from the magnetic quadrupole, due to inelastic collisions, we end with the production of large, deeply degenerate two-component condensates. This technique can be easily extended to other experiments with ultracold mixtures. We point out that our scheme also allows for significant optical access. This provides the possibility to further manipulate the atomic sample by means of engineered optical potentials for future studies on interacting multi-component quantum fluids.

\textbf{Acknowledgments:} We gratefully thank Massimo Inguscio for suggestions and support, Marco Prevedelli and Giacomo Roati for useful discussions, Saverio Bartalini and Julia Emily Bj\o rnst\o m for collaboration at the initial stage of this work. We also acknowledge the continuous help from the technical staff of LENS and the Physics and Astronomy Department of Florence. This work was supported by the European Commission through FP7 Cooperation STREP Project EQuaM (Grant no. 323714) and FET Flagship on Quantum Technologies - Qombs Project (Grant No. 820419). M. M. acknowledges support 
by the Spanish Ministry of Economy, Industry and Competitiveness and the European Regional Development Fund FEDER through Grant No. FIS2015-67161-P (MINECO/FEDER, UE), and by the Basque Government through Grant No. IT986-16.

\bibliography{DoubleBEClast}

\begin{thebibliography}{82}%
\makeatletter
\providecommand \@ifxundefined [1]{%
 \@ifx{#1\undefined}
}%
\providecommand \@ifnum [1]{%
 \ifnum #1\expandafter \@firstoftwo
 \else \expandafter \@secondoftwo
 \fi
}%
\providecommand \@ifx [1]{%
 \ifx #1\expandafter \@firstoftwo
 \else \expandafter \@secondoftwo
 \fi
}%
\providecommand \natexlab [1]{#1}%
\providecommand \enquote  [1]{``#1''}%
\providecommand \bibnamefont  [1]{#1}%
\providecommand \bibfnamefont [1]{#1}%
\providecommand \citenamefont [1]{#1}%
\providecommand \href@noop [0]{\@secondoftwo}%
\providecommand \href [0]{\begingroup \@sanitize@url \@href}%
\providecommand \@href[1]{\@@startlink{#1}\@@href}%
\providecommand \@@href[1]{\endgroup#1\@@endlink}%
\providecommand \@sanitize@url [0]{\catcode `\\12\catcode `\$12\catcode
  `\&12\catcode `\#12\catcode `\^12\catcode `\_12\catcode `\%12\relax}%
\providecommand \@@startlink[1]{}%
\providecommand \@@endlink[0]{}%
\providecommand \url  [0]{\begingroup\@sanitize@url \@url }%
\providecommand \@url [1]{\endgroup\@href {#1}{\urlprefix }}%
\providecommand \urlprefix  [0]{URL }%
\providecommand \Eprint [0]{\href }%
\providecommand \doibase [0]{http://dx.doi.org/}%
\providecommand \selectlanguage [0]{\@gobble}%
\providecommand \bibinfo  [0]{\@secondoftwo}%
\providecommand \bibfield  [0]{\@secondoftwo}%
\providecommand \translation [1]{[#1]}%
\providecommand \BibitemOpen [0]{}%
\providecommand \bibitemStop [0]{}%
\providecommand \bibitemNoStop [0]{.\EOS\space}%
\providecommand \EOS [0]{\spacefactor3000\relax}%
\providecommand \BibitemShut  [1]{\csname bibitem#1\endcsname}%
\let\auto@bib@innerbib\@empty
\bibitem [{\citenamefont {Andreev}\ and\ \citenamefont
  {Bashkin}(1975)}]{AndreevBashkin}%
  \BibitemOpen
  \bibfield  {author} {\bibinfo {author} {\bibfnamefont {A.~F.}\ \bibnamefont
  {Andreev}}\ and\ \bibinfo {author} {\bibfnamefont {E.}~\bibnamefont
  {Bashkin}},\ }\href {http://jetp.ac.ru/cgi-bin/dn/e_042_01_0164.pdf}
  {\bibfield  {journal} {\bibinfo  {journal} {JETP}\ }\textbf {\bibinfo
  {volume} {42}},\ \bibinfo {pages} {164} (\bibinfo {year} {1975})}\BibitemShut
  {NoStop}%
\bibitem [{\citenamefont {Buckley}\ \emph {et~al.}(2004)\citenamefont
  {Buckley}, \citenamefont {Metlitski},\ and\ \citenamefont
  {Zhitnitsky}}]{Buckley}%
  \BibitemOpen
  \bibfield  {author} {\bibinfo {author} {\bibfnamefont {K.~B.~W.}\
  \bibnamefont {Buckley}}, \bibinfo {author} {\bibfnamefont {M.~A.}\
  \bibnamefont {Metlitski}}, \ and\ \bibinfo {author} {\bibfnamefont {A.~R.}\
  \bibnamefont {Zhitnitsky}},\ }\href {\doibase 10.1103/PhysRevLett.92.151102}
  {\bibfield  {journal} {\bibinfo  {journal} {Phys. Rev. Lett.}\ }\textbf
  {\bibinfo {volume} {92}},\ \bibinfo {pages} {151102} (\bibinfo {year}
  {2004})}\BibitemShut {NoStop}%
\bibitem [{\citenamefont {Babaev}\ and\ \citenamefont
  {Speight}(2005)}]{Babaev}%
  \BibitemOpen
  \bibfield  {author} {\bibinfo {author} {\bibfnamefont {E.}~\bibnamefont
  {Babaev}}\ and\ \bibinfo {author} {\bibfnamefont {M.}~\bibnamefont
  {Speight}},\ }\href {\doibase 10.1103/PhysRevB.72.180502} {\bibfield
  {journal} {\bibinfo  {journal} {Phys. Rev. B}\ }\textbf {\bibinfo {volume}
  {72}},\ \bibinfo {pages} {180502} (\bibinfo {year} {2005})}\BibitemShut
  {NoStop}%
\bibitem [{\citenamefont {Anderson}\ \emph {et~al.}(1995)\citenamefont
  {Anderson}, \citenamefont {Ensher}, \citenamefont {Matthews}, \citenamefont
  {Wieman},\ and\ \citenamefont {Cornell}}]{Cornell95}%
  \BibitemOpen
  \bibfield  {author} {\bibinfo {author} {\bibfnamefont {M.~H.}\ \bibnamefont
  {Anderson}}, \bibinfo {author} {\bibfnamefont {J.~R.}\ \bibnamefont
  {Ensher}}, \bibinfo {author} {\bibfnamefont {M.~R.}\ \bibnamefont
  {Matthews}}, \bibinfo {author} {\bibfnamefont {C.~E.}\ \bibnamefont
  {Wieman}}, \ and\ \bibinfo {author} {\bibfnamefont {E.~A.}\ \bibnamefont
  {Cornell}},\ }\href {\doibase 10.1126/science.269.5221.198} {\bibfield
  {journal} {\bibinfo  {journal} {Science}\ }\textbf {\bibinfo {volume}
  {269}},\ \bibinfo {pages} {198} (\bibinfo {year} {1995})}\BibitemShut
  {NoStop}%
\bibitem [{\citenamefont {Davis}\ \emph {et~al.}(1995)\citenamefont {Davis},
  \citenamefont {Mewes}, \citenamefont {Andrews}, \citenamefont {van Druten},
  \citenamefont {Durfee}, \citenamefont {Kurn},\ and\ \citenamefont
  {Ketterle}}]{Daviset95}%
  \BibitemOpen
  \bibfield  {author} {\bibinfo {author} {\bibfnamefont {K.~B.}\ \bibnamefont
  {Davis}}, \bibinfo {author} {\bibfnamefont {M.~O.}\ \bibnamefont {Mewes}},
  \bibinfo {author} {\bibfnamefont {M.~R.}\ \bibnamefont {Andrews}}, \bibinfo
  {author} {\bibfnamefont {N.~J.}\ \bibnamefont {van Druten}}, \bibinfo
  {author} {\bibfnamefont {D.~S.}\ \bibnamefont {Durfee}}, \bibinfo {author}
  {\bibfnamefont {D.~M.}\ \bibnamefont {Kurn}}, \ and\ \bibinfo {author}
  {\bibfnamefont {W.}~\bibnamefont {Ketterle}},\ }\href {\doibase
  10.1103/PhysRevLett.75.3969} {\bibfield  {journal} {\bibinfo  {journal}
  {Phys. Rev. Lett.}\ }\textbf {\bibinfo {volume} {75}},\ \bibinfo {pages}
  {3969} (\bibinfo {year} {1995})}\BibitemShut {NoStop}%
\bibitem [{\citenamefont {Bradley}\ \emph {et~al.}(1995)\citenamefont
  {Bradley}, \citenamefont {Sackett}, \citenamefont {Tollett},\ and\
  \citenamefont {Hulet}}]{Hulet95}%
  \BibitemOpen
  \bibfield  {author} {\bibinfo {author} {\bibfnamefont {C.~C.}\ \bibnamefont
  {Bradley}}, \bibinfo {author} {\bibfnamefont {C.~A.}\ \bibnamefont
  {Sackett}}, \bibinfo {author} {\bibfnamefont {J.~J.}\ \bibnamefont
  {Tollett}}, \ and\ \bibinfo {author} {\bibfnamefont {R.~G.}\ \bibnamefont
  {Hulet}},\ }\href {\doibase 10.1103/PhysRevLett.75.1687} {\bibfield
  {journal} {\bibinfo  {journal} {Phys. Rev. Lett.}\ }\textbf {\bibinfo
  {volume} {75}},\ \bibinfo {pages} {1687} (\bibinfo {year}
  {1995})}\BibitemShut {NoStop}%
\bibitem [{\citenamefont {Myatt}\ \emph {et~al.}(1997)\citenamefont {Myatt},
  \citenamefont {Burt}, \citenamefont {Ghrist}, \citenamefont {Cornell},\ and\
  \citenamefont {Wieman}}]{Cornell97}%
  \BibitemOpen
  \bibfield  {author} {\bibinfo {author} {\bibfnamefont {C.~J.}\ \bibnamefont
  {Myatt}}, \bibinfo {author} {\bibfnamefont {E.~A.}\ \bibnamefont {Burt}},
  \bibinfo {author} {\bibfnamefont {R.~W.}\ \bibnamefont {Ghrist}}, \bibinfo
  {author} {\bibfnamefont {E.~A.}\ \bibnamefont {Cornell}}, \ and\ \bibinfo
  {author} {\bibfnamefont {C.~E.}\ \bibnamefont {Wieman}},\ }\href {\doibase
  10.1103/PhysRevLett.78.586} {\bibfield  {journal} {\bibinfo  {journal} {Phys.
  Rev. Lett.}\ }\textbf {\bibinfo {volume} {78}},\ \bibinfo {pages} {586}
  (\bibinfo {year} {1997})}\BibitemShut {NoStop}%
\bibitem [{\citenamefont {Hall}\ \emph {et~al.}(1998)\citenamefont {Hall},
  \citenamefont {Matthews}, \citenamefont {Wieman},\ and\ \citenamefont
  {Cornell}}]{Cornell98}%
  \BibitemOpen
  \bibfield  {author} {\bibinfo {author} {\bibfnamefont {D.~S.}\ \bibnamefont
  {Hall}}, \bibinfo {author} {\bibfnamefont {M.~R.}\ \bibnamefont {Matthews}},
  \bibinfo {author} {\bibfnamefont {C.~E.}\ \bibnamefont {Wieman}}, \ and\
  \bibinfo {author} {\bibfnamefont {E.~A.}\ \bibnamefont {Cornell}},\ }\href
  {\doibase 10.1103/PhysRevLett.81.1543} {\bibfield  {journal} {\bibinfo
  {journal} {Phys. Rev. Lett.}\ }\textbf {\bibinfo {volume} {81}},\ \bibinfo
  {pages} {1543} (\bibinfo {year} {1998})}\BibitemShut {NoStop}%
\bibitem [{\citenamefont {Maddaloni}\ \emph {et~al.}(2000)\citenamefont
  {Maddaloni}, \citenamefont {Modugno}, \citenamefont {Fort}, \citenamefont
  {Minardi},\ and\ \citenamefont {Inguscio}}]{FortMinardi00}%
  \BibitemOpen
  \bibfield  {author} {\bibinfo {author} {\bibfnamefont {P.}~\bibnamefont
  {Maddaloni}}, \bibinfo {author} {\bibfnamefont {M.}~\bibnamefont {Modugno}},
  \bibinfo {author} {\bibfnamefont {C.}~\bibnamefont {Fort}}, \bibinfo {author}
  {\bibfnamefont {F.}~\bibnamefont {Minardi}}, \ and\ \bibinfo {author}
  {\bibfnamefont {M.}~\bibnamefont {Inguscio}},\ }\href {\doibase
  10.1103/PhysRevLett.85.2413} {\bibfield  {journal} {\bibinfo  {journal}
  {Phys. Rev. Lett.}\ }\textbf {\bibinfo {volume} {85}},\ \bibinfo {pages}
  {2413} (\bibinfo {year} {2000})}\BibitemShut {NoStop}%
\bibitem [{\citenamefont {Truscott}\ \emph {et~al.}(2001)\citenamefont
  {Truscott}, \citenamefont {Strecker}, \citenamefont {McAlexander},
  \citenamefont {Partridge},\ and\ \citenamefont {Hulet}}]{Truscott2001}%
  \BibitemOpen
  \bibfield  {author} {\bibinfo {author} {\bibfnamefont {A.~G.}\ \bibnamefont
  {Truscott}}, \bibinfo {author} {\bibfnamefont {K.~E.}\ \bibnamefont
  {Strecker}}, \bibinfo {author} {\bibfnamefont {W.~I.}\ \bibnamefont
  {McAlexander}}, \bibinfo {author} {\bibfnamefont {G.~B.}\ \bibnamefont
  {Partridge}}, \ and\ \bibinfo {author} {\bibfnamefont {R.~G.}\ \bibnamefont
  {Hulet}},\ }\href {\doibase 10.1126/science.1059318} {\bibfield  {journal}
  {\bibinfo  {journal} {Science}\ }\textbf {\bibinfo {volume} {291}},\ \bibinfo
  {pages} {2570} (\bibinfo {year} {2001})},\ \Eprint
  {http://arxiv.org/abs/http://science.sciencemag.org/content/291/5513/2570.full.pdf}
  {http://science.sciencemag.org/content/291/5513/2570.full.pdf} \BibitemShut
  {NoStop}%
\bibitem [{\citenamefont {Schreck}\ \emph {et~al.}(2001)\citenamefont
  {Schreck}, \citenamefont {Khaykovich}, \citenamefont {Corwin}, \citenamefont
  {Ferrari}, \citenamefont {Bourdel}, \citenamefont {Cubizolles},\ and\
  \citenamefont {Salomon}}]{Schreck2001}%
  \BibitemOpen
  \bibfield  {author} {\bibinfo {author} {\bibfnamefont {F.}~\bibnamefont
  {Schreck}}, \bibinfo {author} {\bibfnamefont {L.}~\bibnamefont {Khaykovich}},
  \bibinfo {author} {\bibfnamefont {K.~L.}\ \bibnamefont {Corwin}}, \bibinfo
  {author} {\bibfnamefont {G.}~\bibnamefont {Ferrari}}, \bibinfo {author}
  {\bibfnamefont {T.}~\bibnamefont {Bourdel}}, \bibinfo {author} {\bibfnamefont
  {J.}~\bibnamefont {Cubizolles}}, \ and\ \bibinfo {author} {\bibfnamefont
  {C.}~\bibnamefont {Salomon}},\ }\href {\doibase
  10.1103/PhysRevLett.87.080403} {\bibfield  {journal} {\bibinfo  {journal}
  {Phys. Rev. Lett.}\ }\textbf {\bibinfo {volume} {87}},\ \bibinfo {pages}
  {080403} (\bibinfo {year} {2001})}\BibitemShut {NoStop}%
\bibitem [{\citenamefont {McNamara}\ \emph {et~al.}(2006)\citenamefont
  {McNamara}, \citenamefont {Jeltes}, \citenamefont {Tychkov}, \citenamefont
  {Hogervorst},\ and\ \citenamefont {Vassen}}]{McNamara2006}%
  \BibitemOpen
  \bibfield  {author} {\bibinfo {author} {\bibfnamefont {J.~M.}\ \bibnamefont
  {McNamara}}, \bibinfo {author} {\bibfnamefont {T.}~\bibnamefont {Jeltes}},
  \bibinfo {author} {\bibfnamefont {A.~S.}\ \bibnamefont {Tychkov}}, \bibinfo
  {author} {\bibfnamefont {W.}~\bibnamefont {Hogervorst}}, \ and\ \bibinfo
  {author} {\bibfnamefont {W.}~\bibnamefont {Vassen}},\ }\href {\doibase
  10.1103/PhysRevLett.97.080404} {\bibfield  {journal} {\bibinfo  {journal}
  {Phys. Rev. Lett.}\ }\textbf {\bibinfo {volume} {97}},\ \bibinfo {pages}
  {080404} (\bibinfo {year} {2006})}\BibitemShut {NoStop}%
\bibitem [{\citenamefont {Papp}\ \emph {et~al.}(2008)\citenamefont {Papp},
  \citenamefont {Pino},\ and\ \citenamefont {Wieman}}]{Wieman2008}%
  \BibitemOpen
  \bibfield  {author} {\bibinfo {author} {\bibfnamefont {S.~B.}\ \bibnamefont
  {Papp}}, \bibinfo {author} {\bibfnamefont {J.~M.}\ \bibnamefont {Pino}}, \
  and\ \bibinfo {author} {\bibfnamefont {C.~E.}\ \bibnamefont {Wieman}},\
  }\href {\doibase 10.1103/PhysRevLett.101.040402} {\bibfield  {journal}
  {\bibinfo  {journal} {Phys. Rev. Lett.}\ }\textbf {\bibinfo {volume} {101}},\
  \bibinfo {pages} {040402} (\bibinfo {year} {2008})}\BibitemShut {NoStop}%
\bibitem [{\citenamefont {Tey}\ \emph {et~al.}(2010)\citenamefont {Tey},
  \citenamefont {Stellmer}, \citenamefont {Grimm},\ and\ \citenamefont
  {Schreck}}]{Tey2010}%
  \BibitemOpen
  \bibfield  {author} {\bibinfo {author} {\bibfnamefont {M.~K.}\ \bibnamefont
  {Tey}}, \bibinfo {author} {\bibfnamefont {S.}~\bibnamefont {Stellmer}},
  \bibinfo {author} {\bibfnamefont {R.}~\bibnamefont {Grimm}}, \ and\ \bibinfo
  {author} {\bibfnamefont {F.}~\bibnamefont {Schreck}},\ }\href {\doibase
  10.1103/PhysRevA.82.011608} {\bibfield  {journal} {\bibinfo  {journal} {Phys.
  Rev. A}\ }\textbf {\bibinfo {volume} {82}},\ \bibinfo {pages} {011608}
  (\bibinfo {year} {2010})}\BibitemShut {NoStop}%
\bibitem [{\citenamefont {Lu}\ \emph {et~al.}(2011)\citenamefont {Lu},
  \citenamefont {Burdick}, \citenamefont {Youn},\ and\ \citenamefont
  {Lev}}]{Lu2011}%
  \BibitemOpen
  \bibfield  {author} {\bibinfo {author} {\bibfnamefont {M.}~\bibnamefont
  {Lu}}, \bibinfo {author} {\bibfnamefont {N.~Q.}\ \bibnamefont {Burdick}},
  \bibinfo {author} {\bibfnamefont {S.~H.}\ \bibnamefont {Youn}}, \ and\
  \bibinfo {author} {\bibfnamefont {B.~L.}\ \bibnamefont {Lev}},\ }\href
  {\doibase 10.1103/PhysRevLett.107.190401} {\bibfield  {journal} {\bibinfo
  {journal} {Phys. Rev. Lett.}\ }\textbf {\bibinfo {volume} {107}},\ \bibinfo
  {pages} {190401} (\bibinfo {year} {2011})}\BibitemShut {NoStop}%
\bibitem [{\citenamefont {Sugawa}\ \emph {et~al.}(2011)\citenamefont {Sugawa},
  \citenamefont {Yamazaki}, \citenamefont {Taie},\ and\ \citenamefont
  {Takahashi}}]{Sugawa2011}%
  \BibitemOpen
  \bibfield  {author} {\bibinfo {author} {\bibfnamefont {S.}~\bibnamefont
  {Sugawa}}, \bibinfo {author} {\bibfnamefont {R.}~\bibnamefont {Yamazaki}},
  \bibinfo {author} {\bibfnamefont {S.}~\bibnamefont {Taie}}, \ and\ \bibinfo
  {author} {\bibfnamefont {Y.}~\bibnamefont {Takahashi}},\ }\href {\doibase
  10.1103/PhysRevA.84.011610} {\bibfield  {journal} {\bibinfo  {journal} {Phys.
  Rev. A}\ }\textbf {\bibinfo {volume} {84}},\ \bibinfo {pages} {011610}
  (\bibinfo {year} {2011})}\BibitemShut {NoStop}%
\bibitem [{\citenamefont {Stellmer}\ \emph {et~al.}(2013)\citenamefont
  {Stellmer}, \citenamefont {Grimm},\ and\ \citenamefont
  {Schreck}}]{Grimm2013}%
  \BibitemOpen
  \bibfield  {author} {\bibinfo {author} {\bibfnamefont {S.}~\bibnamefont
  {Stellmer}}, \bibinfo {author} {\bibfnamefont {R.}~\bibnamefont {Grimm}}, \
  and\ \bibinfo {author} {\bibfnamefont {F.}~\bibnamefont {Schreck}},\ }\href
  {\doibase 10.1103/PhysRevA.87.013611} {\bibfield  {journal} {\bibinfo
  {journal} {Phys. Rev. A}\ }\textbf {\bibinfo {volume} {87}},\ \bibinfo
  {pages} {013611} (\bibinfo {year} {2013})}\BibitemShut {NoStop}%
\bibitem [{\citenamefont {Ferrier-Barbut}\ \emph {et~al.}(2014)\citenamefont
  {Ferrier-Barbut}, \citenamefont {Delehaye}, \citenamefont {Laurent},
  \citenamefont {Grier}, \citenamefont {Pierce}, \citenamefont {Rem},
  \citenamefont {Chevy},\ and\ \citenamefont {Salomon}}]{Salomon2014}%
  \BibitemOpen
  \bibfield  {author} {\bibinfo {author} {\bibfnamefont {I.}~\bibnamefont
  {Ferrier-Barbut}}, \bibinfo {author} {\bibfnamefont {M.}~\bibnamefont
  {Delehaye}}, \bibinfo {author} {\bibfnamefont {S.}~\bibnamefont {Laurent}},
  \bibinfo {author} {\bibfnamefont {A.~T.}\ \bibnamefont {Grier}}, \bibinfo
  {author} {\bibfnamefont {M.}~\bibnamefont {Pierce}}, \bibinfo {author}
  {\bibfnamefont {B.~S.}\ \bibnamefont {Rem}}, \bibinfo {author} {\bibfnamefont
  {F.}~\bibnamefont {Chevy}}, \ and\ \bibinfo {author} {\bibfnamefont
  {C.}~\bibnamefont {Salomon}},\ }\href {\doibase 10.1126/science.1255380}
  {\bibfield  {journal} {\bibinfo  {journal} {Science}\ }\textbf {\bibinfo
  {volume} {345}},\ \bibinfo {pages} {1035} (\bibinfo {year}
  {2014})}\BibitemShut {NoStop}%
\bibitem [{\citenamefont {Modugno}\ \emph {et~al.}(2002)\citenamefont
  {Modugno}, \citenamefont {Modugno}, \citenamefont {Riboli}, \citenamefont
  {Roati},\ and\ \citenamefont {Inguscio}}]{Modugno2002}%
  \BibitemOpen
  \bibfield  {author} {\bibinfo {author} {\bibfnamefont {G.}~\bibnamefont
  {Modugno}}, \bibinfo {author} {\bibfnamefont {M.}~\bibnamefont {Modugno}},
  \bibinfo {author} {\bibfnamefont {F.}~\bibnamefont {Riboli}}, \bibinfo
  {author} {\bibfnamefont {G.}~\bibnamefont {Roati}}, \ and\ \bibinfo {author}
  {\bibfnamefont {M.}~\bibnamefont {Inguscio}},\ }\href {\doibase
  10.1103/PhysRevLett.89.190404} {\bibfield  {journal} {\bibinfo  {journal}
  {Phys. Rev. Lett.}\ }\textbf {\bibinfo {volume} {89}},\ \bibinfo {pages}
  {190404} (\bibinfo {year} {2002})}\BibitemShut {NoStop}%
\bibitem [{\citenamefont {Hadzibabic}\ \emph {et~al.}(2002)\citenamefont
  {Hadzibabic}, \citenamefont {Stan}, \citenamefont {Dieckmann}, \citenamefont
  {Gupta}, \citenamefont {Zwierlein}, \citenamefont {G\"orlitz},\ and\
  \citenamefont {Ketterle}}]{Hadzibabic2002}%
  \BibitemOpen
  \bibfield  {author} {\bibinfo {author} {\bibfnamefont {Z.}~\bibnamefont
  {Hadzibabic}}, \bibinfo {author} {\bibfnamefont {C.~A.}\ \bibnamefont
  {Stan}}, \bibinfo {author} {\bibfnamefont {K.}~\bibnamefont {Dieckmann}},
  \bibinfo {author} {\bibfnamefont {S.}~\bibnamefont {Gupta}}, \bibinfo
  {author} {\bibfnamefont {M.~W.}\ \bibnamefont {Zwierlein}}, \bibinfo {author}
  {\bibfnamefont {A.}~\bibnamefont {G\"orlitz}}, \ and\ \bibinfo {author}
  {\bibfnamefont {W.}~\bibnamefont {Ketterle}},\ }\href {\doibase
  10.1103/PhysRevLett.88.160401} {\bibfield  {journal} {\bibinfo  {journal}
  {Phys. Rev. Lett.}\ }\textbf {\bibinfo {volume} {88}},\ \bibinfo {pages}
  {160401} (\bibinfo {year} {2002})}\BibitemShut {NoStop}%
\bibitem [{\citenamefont {Roati}\ \emph {et~al.}(2002)\citenamefont {Roati},
  \citenamefont {Riboli}, \citenamefont {Modugno},\ and\ \citenamefont
  {Inguscio}}]{Roati2002}%
  \BibitemOpen
  \bibfield  {author} {\bibinfo {author} {\bibfnamefont {G.}~\bibnamefont
  {Roati}}, \bibinfo {author} {\bibfnamefont {F.}~\bibnamefont {Riboli}},
  \bibinfo {author} {\bibfnamefont {G.}~\bibnamefont {Modugno}}, \ and\
  \bibinfo {author} {\bibfnamefont {M.}~\bibnamefont {Inguscio}},\ }\href
  {\doibase 10.1103/PhysRevLett.89.150403} {\bibfield  {journal} {\bibinfo
  {journal} {Phys. Rev. Lett.}\ }\textbf {\bibinfo {volume} {89}},\ \bibinfo
  {pages} {150403} (\bibinfo {year} {2002})}\BibitemShut {NoStop}%
\bibitem [{\citenamefont {Silber}\ \emph {et~al.}(2005)\citenamefont {Silber},
  \citenamefont {G\"unther}, \citenamefont {Marzok}, \citenamefont {Deh},
  \citenamefont {Courteille},\ and\ \citenamefont
  {Zimmermann}}]{Zimmermann2005}%
  \BibitemOpen
  \bibfield  {author} {\bibinfo {author} {\bibfnamefont {C.}~\bibnamefont
  {Silber}}, \bibinfo {author} {\bibfnamefont {S.}~\bibnamefont {G\"unther}},
  \bibinfo {author} {\bibfnamefont {C.}~\bibnamefont {Marzok}}, \bibinfo
  {author} {\bibfnamefont {B.}~\bibnamefont {Deh}}, \bibinfo {author}
  {\bibfnamefont {P.~W.}\ \bibnamefont {Courteille}}, \ and\ \bibinfo {author}
  {\bibfnamefont {C.}~\bibnamefont {Zimmermann}},\ }\href {\doibase
  10.1103/PhysRevLett.95.170408} {\bibfield  {journal} {\bibinfo  {journal}
  {Phys. Rev. Lett.}\ }\textbf {\bibinfo {volume} {95}},\ \bibinfo {pages}
  {170408} (\bibinfo {year} {2005})}\BibitemShut {NoStop}%
\bibitem [{\citenamefont {Taglieber}\ \emph {et~al.}(2008)\citenamefont
  {Taglieber}, \citenamefont {Voigt}, \citenamefont {Aoki}, \citenamefont
  {H\"ansch},\ and\ \citenamefont {Dieckmann}}]{Dieckmann2008}%
  \BibitemOpen
  \bibfield  {author} {\bibinfo {author} {\bibfnamefont {M.}~\bibnamefont
  {Taglieber}}, \bibinfo {author} {\bibfnamefont {A.-C.}\ \bibnamefont
  {Voigt}}, \bibinfo {author} {\bibfnamefont {T.}~\bibnamefont {Aoki}},
  \bibinfo {author} {\bibfnamefont {T.~W.}\ \bibnamefont {H\"ansch}}, \ and\
  \bibinfo {author} {\bibfnamefont {K.}~\bibnamefont {Dieckmann}},\ }\href
  {\doibase 10.1103/PhysRevLett.100.010401} {\bibfield  {journal} {\bibinfo
  {journal} {Phys. Rev. Lett.}\ }\textbf {\bibinfo {volume} {100}},\ \bibinfo
  {pages} {010401} (\bibinfo {year} {2008})}\BibitemShut {NoStop}%
\bibitem [{\citenamefont {Thalhammer}\ \emph {et~al.}(2008)\citenamefont
  {Thalhammer}, \citenamefont {Barontini}, \citenamefont {De~Sarlo},
  \citenamefont {Catani}, \citenamefont {Minardi},\ and\ \citenamefont
  {Inguscio}}]{Thalhammer2008}%
  \BibitemOpen
  \bibfield  {author} {\bibinfo {author} {\bibfnamefont {G.}~\bibnamefont
  {Thalhammer}}, \bibinfo {author} {\bibfnamefont {G.}~\bibnamefont
  {Barontini}}, \bibinfo {author} {\bibfnamefont {L.}~\bibnamefont {De~Sarlo}},
  \bibinfo {author} {\bibfnamefont {J.}~\bibnamefont {Catani}}, \bibinfo
  {author} {\bibfnamefont {F.}~\bibnamefont {Minardi}}, \ and\ \bibinfo
  {author} {\bibfnamefont {M.}~\bibnamefont {Inguscio}},\ }\href {\doibase
  10.1103/PhysRevLett.100.210402} {\bibfield  {journal} {\bibinfo  {journal}
  {Phys. Rev. Lett.}\ }\textbf {\bibinfo {volume} {100}},\ \bibinfo {pages}
  {210402} (\bibinfo {year} {2008})}\BibitemShut {NoStop}%
\bibitem [{\citenamefont {Lercher}\ \emph {et~al.}(2011)\citenamefont
  {Lercher}, \citenamefont {Takekoshi}, \citenamefont {Debatin}, \citenamefont
  {Schuster}, \citenamefont {Rameshan}, \citenamefont {Ferlaino}, \citenamefont
  {Grimm},\ and\ \citenamefont {N{\"a}gerl}}]{Lercher2011}%
  \BibitemOpen
  \bibfield  {author} {\bibinfo {author} {\bibfnamefont {A.~D.}\ \bibnamefont
  {Lercher}}, \bibinfo {author} {\bibfnamefont {T.}~\bibnamefont {Takekoshi}},
  \bibinfo {author} {\bibfnamefont {M.}~\bibnamefont {Debatin}}, \bibinfo
  {author} {\bibfnamefont {B.}~\bibnamefont {Schuster}}, \bibinfo {author}
  {\bibfnamefont {R.}~\bibnamefont {Rameshan}}, \bibinfo {author}
  {\bibfnamefont {F.}~\bibnamefont {Ferlaino}}, \bibinfo {author}
  {\bibfnamefont {R.}~\bibnamefont {Grimm}}, \ and\ \bibinfo {author}
  {\bibfnamefont {H.~C.}\ \bibnamefont {N{\"a}gerl}},\ }\href {\doibase
  10.1140/epjd/e2011-20015-6} {\bibfield  {journal} {\bibinfo  {journal} {The
  European Physical Journal D}\ }\textbf {\bibinfo {volume} {65}},\ \bibinfo
  {pages} {3} (\bibinfo {year} {2011})}\BibitemShut {NoStop}%
\bibitem [{\citenamefont {McCarron}\ \emph {et~al.}(2011)\citenamefont
  {McCarron}, \citenamefont {Cho}, \citenamefont {Jenkin}, \citenamefont
  {K\"oppinger},\ and\ \citenamefont {Cornish}}]{Cornish2011}%
  \BibitemOpen
  \bibfield  {author} {\bibinfo {author} {\bibfnamefont {D.~J.}\ \bibnamefont
  {McCarron}}, \bibinfo {author} {\bibfnamefont {H.~W.}\ \bibnamefont {Cho}},
  \bibinfo {author} {\bibfnamefont {D.~L.}\ \bibnamefont {Jenkin}}, \bibinfo
  {author} {\bibfnamefont {M.~P.}\ \bibnamefont {K\"oppinger}}, \ and\ \bibinfo
  {author} {\bibfnamefont {S.~L.}\ \bibnamefont {Cornish}},\ }\href {\doibase
  10.1103/PhysRevA.84.011603} {\bibfield  {journal} {\bibinfo  {journal} {Phys.
  Rev. A}\ }\textbf {\bibinfo {volume} {84}},\ \bibinfo {pages} {011603}
  (\bibinfo {year} {2011})}\BibitemShut {NoStop}%
\bibitem [{\citenamefont {Park}\ \emph {et~al.}(2012)\citenamefont {Park},
  \citenamefont {Wu}, \citenamefont {Santiago}, \citenamefont {Tiecke},
  \citenamefont {Will}, \citenamefont {Ahmadi},\ and\ \citenamefont
  {Zwierlein}}]{Park2012}%
  \BibitemOpen
  \bibfield  {author} {\bibinfo {author} {\bibfnamefont {J.~W.}\ \bibnamefont
  {Park}}, \bibinfo {author} {\bibfnamefont {C.-H.}\ \bibnamefont {Wu}},
  \bibinfo {author} {\bibfnamefont {I.}~\bibnamefont {Santiago}}, \bibinfo
  {author} {\bibfnamefont {T.~G.}\ \bibnamefont {Tiecke}}, \bibinfo {author}
  {\bibfnamefont {S.}~\bibnamefont {Will}}, \bibinfo {author} {\bibfnamefont
  {P.}~\bibnamefont {Ahmadi}}, \ and\ \bibinfo {author} {\bibfnamefont {M.~W.}\
  \bibnamefont {Zwierlein}},\ }\href {\doibase 10.1103/PhysRevA.85.051602}
  {\bibfield  {journal} {\bibinfo  {journal} {Phys. Rev. A}\ }\textbf {\bibinfo
  {volume} {85}},\ \bibinfo {pages} {051602} (\bibinfo {year}
  {2012})}\BibitemShut {NoStop}%
\bibitem [{\citenamefont {Repp}\ \emph {et~al.}(2013)\citenamefont {Repp},
  \citenamefont {Pires}, \citenamefont {Ulmanis}, \citenamefont {Heck},
  \citenamefont {Kuhnle}, \citenamefont {Weidem\"uller},\ and\ \citenamefont
  {Tiemann}}]{Repp2013}%
  \BibitemOpen
  \bibfield  {author} {\bibinfo {author} {\bibfnamefont {M.}~\bibnamefont
  {Repp}}, \bibinfo {author} {\bibfnamefont {R.}~\bibnamefont {Pires}},
  \bibinfo {author} {\bibfnamefont {J.}~\bibnamefont {Ulmanis}}, \bibinfo
  {author} {\bibfnamefont {R.}~\bibnamefont {Heck}}, \bibinfo {author}
  {\bibfnamefont {E.~D.}\ \bibnamefont {Kuhnle}}, \bibinfo {author}
  {\bibfnamefont {M.}~\bibnamefont {Weidem\"uller}}, \ and\ \bibinfo {author}
  {\bibfnamefont {E.}~\bibnamefont {Tiemann}},\ }\href {\doibase
  10.1103/PhysRevA.87.010701} {\bibfield  {journal} {\bibinfo  {journal} {Phys.
  Rev. A}\ }\textbf {\bibinfo {volume} {87}},\ \bibinfo {pages} {010701}
  (\bibinfo {year} {2013})}\BibitemShut {NoStop}%
\bibitem [{\citenamefont {Pasquiou}\ \emph {et~al.}(2013)\citenamefont
  {Pasquiou}, \citenamefont {Bayerle}, \citenamefont {Tzanova}, \citenamefont
  {Stellmer}, \citenamefont {Szczepkowski}, \citenamefont {Parigger},
  \citenamefont {Grimm},\ and\ \citenamefont {Schreck}}]{Schreck2013}%
  \BibitemOpen
  \bibfield  {author} {\bibinfo {author} {\bibfnamefont {B.}~\bibnamefont
  {Pasquiou}}, \bibinfo {author} {\bibfnamefont {A.}~\bibnamefont {Bayerle}},
  \bibinfo {author} {\bibfnamefont {S.~M.}\ \bibnamefont {Tzanova}}, \bibinfo
  {author} {\bibfnamefont {S.}~\bibnamefont {Stellmer}}, \bibinfo {author}
  {\bibfnamefont {J.}~\bibnamefont {Szczepkowski}}, \bibinfo {author}
  {\bibfnamefont {M.}~\bibnamefont {Parigger}}, \bibinfo {author}
  {\bibfnamefont {R.}~\bibnamefont {Grimm}}, \ and\ \bibinfo {author}
  {\bibfnamefont {F.}~\bibnamefont {Schreck}},\ }\href {\doibase
  10.1103/PhysRevA.88.023601} {\bibfield  {journal} {\bibinfo  {journal} {Phys.
  Rev. A}\ }\textbf {\bibinfo {volume} {88}},\ \bibinfo {pages} {023601}
  (\bibinfo {year} {2013})}\BibitemShut {NoStop}%
\bibitem [{\citenamefont {Wacker}\ \emph {et~al.}(2015)\citenamefont {Wacker},
  \citenamefont {J\o{}rgensen}, \citenamefont {Birkmose}, \citenamefont
  {Horchani}, \citenamefont {Ertmer}, \citenamefont {Klempt}, \citenamefont
  {Winter}, \citenamefont {Sherson},\ and\ \citenamefont
  {Arlt}}]{wacker_tunable_2015}%
  \BibitemOpen
  \bibfield  {author} {\bibinfo {author} {\bibfnamefont {L.}~\bibnamefont
  {Wacker}}, \bibinfo {author} {\bibfnamefont {N.~B.}\ \bibnamefont
  {J\o{}rgensen}}, \bibinfo {author} {\bibfnamefont {D.}~\bibnamefont
  {Birkmose}}, \bibinfo {author} {\bibfnamefont {R.}~\bibnamefont {Horchani}},
  \bibinfo {author} {\bibfnamefont {W.}~\bibnamefont {Ertmer}}, \bibinfo
  {author} {\bibfnamefont {C.}~\bibnamefont {Klempt}}, \bibinfo {author}
  {\bibfnamefont {N.}~\bibnamefont {Winter}}, \bibinfo {author} {\bibfnamefont
  {J.}~\bibnamefont {Sherson}}, \ and\ \bibinfo {author} {\bibfnamefont
  {J.~J.}\ \bibnamefont {Arlt}},\ }\href {\doibase 10.1103/PhysRevA.92.053602}
  {\bibfield  {journal} {\bibinfo  {journal} {Phys. Rev. A}\ }\textbf {\bibinfo
  {volume} {92}},\ \bibinfo {pages} {053602} (\bibinfo {year}
  {2015})}\BibitemShut {NoStop}%
\bibitem [{\citenamefont {Wang}\ \emph {et~al.}(2016)\citenamefont {Wang},
  \citenamefont {Li}, \citenamefont {Xiong},\ and\ \citenamefont
  {Wang}}]{Wang2016}%
  \BibitemOpen
  \bibfield  {author} {\bibinfo {author} {\bibfnamefont {F.}~\bibnamefont
  {Wang}}, \bibinfo {author} {\bibfnamefont {X.}~\bibnamefont {Li}}, \bibinfo
  {author} {\bibfnamefont {D.}~\bibnamefont {Xiong}}, \ and\ \bibinfo {author}
  {\bibfnamefont {D.}~\bibnamefont {Wang}},\ }\href@noop {} {\bibfield
  {journal} {\bibinfo  {journal} {J. Phys. B}\ }\textbf {\bibinfo {volume}
  {49}},\ \bibinfo {pages} {015302} (\bibinfo {year} {2016})}\BibitemShut
  {NoStop}%
\bibitem [{\citenamefont {Roy}\ \emph {et~al.}(2017)\citenamefont {Roy},
  \citenamefont {Green}, \citenamefont {Bowler},\ and\ \citenamefont
  {Gupta}}]{Roy2017}%
  \BibitemOpen
  \bibfield  {author} {\bibinfo {author} {\bibfnamefont {R.}~\bibnamefont
  {Roy}}, \bibinfo {author} {\bibfnamefont {A.}~\bibnamefont {Green}}, \bibinfo
  {author} {\bibfnamefont {R.}~\bibnamefont {Bowler}}, \ and\ \bibinfo {author}
  {\bibfnamefont {S.}~\bibnamefont {Gupta}},\ }\href {\doibase
  10.1103/PhysRevLett.118.055301} {\bibfield  {journal} {\bibinfo  {journal}
  {Phys. Rev. Lett.}\ }\textbf {\bibinfo {volume} {118}},\ \bibinfo {pages}
  {055301} (\bibinfo {year} {2017})}\BibitemShut {NoStop}%
\bibitem [{\citenamefont {Schulze}\ \emph {et~al.}(2018)\citenamefont
  {Schulze}, \citenamefont {Hartmann}, \citenamefont {Voges}, \citenamefont
  {Gempel}, \citenamefont {Tiemann}, \citenamefont {Zenesini},\ and\
  \citenamefont {Ospelkaus}}]{Ospelkaus2018}%
  \BibitemOpen
  \bibfield  {author} {\bibinfo {author} {\bibfnamefont {T.~A.}\ \bibnamefont
  {Schulze}}, \bibinfo {author} {\bibfnamefont {T.}~\bibnamefont {Hartmann}},
  \bibinfo {author} {\bibfnamefont {K.~K.}\ \bibnamefont {Voges}}, \bibinfo
  {author} {\bibfnamefont {M.~W.}\ \bibnamefont {Gempel}}, \bibinfo {author}
  {\bibfnamefont {E.}~\bibnamefont {Tiemann}}, \bibinfo {author} {\bibfnamefont
  {A.}~\bibnamefont {Zenesini}}, \ and\ \bibinfo {author} {\bibfnamefont
  {S.}~\bibnamefont {Ospelkaus}},\ }\href {\doibase 10.1103/PhysRevA.97.023623}
  {\bibfield  {journal} {\bibinfo  {journal} {Phys. Rev. A}\ }\textbf {\bibinfo
  {volume} {97}},\ \bibinfo {pages} {023623} (\bibinfo {year}
  {2018})}\BibitemShut {NoStop}%
\bibitem [{\citenamefont {{Trautmann}}\ \emph {et~al.}(2018)\citenamefont
  {{Trautmann}}, \citenamefont {{Ilzh{\"o}fer}}, \citenamefont {{Durastante}},
  \citenamefont {{Politi}}, \citenamefont {{Sohmen}}, \citenamefont {{Mark}},\
  and\ \citenamefont {{Ferlaino}}}]{Ferlaino2018}%
  \BibitemOpen
  \bibfield  {author} {\bibinfo {author} {\bibfnamefont {A.}~\bibnamefont
  {{Trautmann}}}, \bibinfo {author} {\bibfnamefont {P.}~\bibnamefont
  {{Ilzh{\"o}fer}}}, \bibinfo {author} {\bibfnamefont {G.}~\bibnamefont
  {{Durastante}}}, \bibinfo {author} {\bibfnamefont {C.}~\bibnamefont
  {{Politi}}}, \bibinfo {author} {\bibfnamefont {M.}~\bibnamefont {{Sohmen}}},
  \bibinfo {author} {\bibfnamefont {M.~J.}\ \bibnamefont {{Mark}}}, \ and\
  \bibinfo {author} {\bibfnamefont {F.}~\bibnamefont {{Ferlaino}}},\
  }\href@noop {} {\bibfield  {journal} {\bibinfo  {journal} {ArXiv e-prints}\ }
  (\bibinfo {year} {2018})},\ \Eprint {http://arxiv.org/abs/1807.07555}
  {arXiv:1807.07555 [cond-mat.quant-gas]} \BibitemShut {NoStop}%
\bibitem [{\citenamefont {Schweikhard}\ \emph {et~al.}(2004)\citenamefont
  {Schweikhard}, \citenamefont {Coddington}, \citenamefont {Engels},
  \citenamefont {Tung},\ and\ \citenamefont {Cornell}}]{Cornell2004}%
  \BibitemOpen
  \bibfield  {author} {\bibinfo {author} {\bibfnamefont {V.}~\bibnamefont
  {Schweikhard}}, \bibinfo {author} {\bibfnamefont {I.}~\bibnamefont
  {Coddington}}, \bibinfo {author} {\bibfnamefont {P.}~\bibnamefont {Engels}},
  \bibinfo {author} {\bibfnamefont {S.}~\bibnamefont {Tung}}, \ and\ \bibinfo
  {author} {\bibfnamefont {E.~A.}\ \bibnamefont {Cornell}},\ }\href {\doibase
  10.1103/PhysRevLett.93.210403} {\bibfield  {journal} {\bibinfo  {journal}
  {Phys. Rev. Lett.}\ }\textbf {\bibinfo {volume} {93}},\ \bibinfo {pages}
  {210403} (\bibinfo {year} {2004})}\BibitemShut {NoStop}%
\bibitem [{\citenamefont {Hamner}\ \emph {et~al.}(2011)\citenamefont {Hamner},
  \citenamefont {Chang}, \citenamefont {Engels},\ and\ \citenamefont
  {Hoefer}}]{Hoefer2011}%
  \BibitemOpen
  \bibfield  {author} {\bibinfo {author} {\bibfnamefont {C.}~\bibnamefont
  {Hamner}}, \bibinfo {author} {\bibfnamefont {J.~J.}\ \bibnamefont {Chang}},
  \bibinfo {author} {\bibfnamefont {P.}~\bibnamefont {Engels}}, \ and\ \bibinfo
  {author} {\bibfnamefont {M.~A.}\ \bibnamefont {Hoefer}},\ }\href {\doibase
  10.1103/PhysRevLett.106.065302} {\bibfield  {journal} {\bibinfo  {journal}
  {Phys. Rev. Lett.}\ }\textbf {\bibinfo {volume} {106}},\ \bibinfo {pages}
  {065302} (\bibinfo {year} {2011})}\BibitemShut {NoStop}%
\bibitem [{\citenamefont {Yao}\ \emph {et~al.}(2016)\citenamefont {Yao},
  \citenamefont {Chen}, \citenamefont {Wu}, \citenamefont {Liu}, \citenamefont
  {Wang}, \citenamefont {Jiang}, \citenamefont {Deng}, \citenamefont {Chen},\
  and\ \citenamefont {Pan}}]{Yao2016}%
  \BibitemOpen
  \bibfield  {author} {\bibinfo {author} {\bibfnamefont {X.-C.}\ \bibnamefont
  {Yao}}, \bibinfo {author} {\bibfnamefont {H.-Z.}\ \bibnamefont {Chen}},
  \bibinfo {author} {\bibfnamefont {Y.-P.}\ \bibnamefont {Wu}}, \bibinfo
  {author} {\bibfnamefont {X.-P.}\ \bibnamefont {Liu}}, \bibinfo {author}
  {\bibfnamefont {X.-Q.}\ \bibnamefont {Wang}}, \bibinfo {author}
  {\bibfnamefont {X.}~\bibnamefont {Jiang}}, \bibinfo {author} {\bibfnamefont
  {Y.}~\bibnamefont {Deng}}, \bibinfo {author} {\bibfnamefont {Y.-A.}\
  \bibnamefont {Chen}}, \ and\ \bibinfo {author} {\bibfnamefont {J.-W.}\
  \bibnamefont {Pan}},\ }\href {\doibase 10.1103/PhysRevLett.117.145301}
  {\bibfield  {journal} {\bibinfo  {journal} {Phys. Rev. Lett.}\ }\textbf
  {\bibinfo {volume} {117}},\ \bibinfo {pages} {145301} (\bibinfo {year}
  {2016})}\BibitemShut {NoStop}%
\bibitem [{\citenamefont {Lee}\ \emph {et~al.}(2018)\citenamefont {Lee},
  \citenamefont {Jørgensen}, \citenamefont {Wacker}, \citenamefont {Skou},
  \citenamefont {Skalmstang}, \citenamefont {Arlt},\ and\ \citenamefont
  {Proukakis}}]{Proukakis2018}%
  \BibitemOpen
  \bibfield  {author} {\bibinfo {author} {\bibfnamefont {K.~L.}\ \bibnamefont
  {Lee}}, \bibinfo {author} {\bibfnamefont {N.~B.}\ \bibnamefont {Jørgensen}},
  \bibinfo {author} {\bibfnamefont {L.~J.}\ \bibnamefont {Wacker}}, \bibinfo
  {author} {\bibfnamefont {M.~G.}\ \bibnamefont {Skou}}, \bibinfo {author}
  {\bibfnamefont {K.~T.}\ \bibnamefont {Skalmstang}}, \bibinfo {author}
  {\bibfnamefont {J.~J.}\ \bibnamefont {Arlt}}, \ and\ \bibinfo {author}
  {\bibfnamefont {N.~P.}\ \bibnamefont {Proukakis}},\ }\href
  {http://stacks.iop.org/1367-2630/20/i=5/a=053004} {\bibfield  {journal}
  {\bibinfo  {journal} {New J. Phys.}\ }\textbf {\bibinfo {volume} {20}},\
  \bibinfo {pages} {053004} (\bibinfo {year} {2018})}\BibitemShut {NoStop}%
\bibitem [{\citenamefont {Kuklov}\ and\ \citenamefont
  {Svistunov}(2003)}]{kuklov_counterflow_2003}%
  \BibitemOpen
  \bibfield  {author} {\bibinfo {author} {\bibfnamefont {A.~B.}\ \bibnamefont
  {Kuklov}}\ and\ \bibinfo {author} {\bibfnamefont {B.~V.}\ \bibnamefont
  {Svistunov}},\ }\href {\doibase 10.1103/PhysRevLett.90.100401} {\bibfield
  {journal} {\bibinfo  {journal} {Physical Review Letters}\ }\textbf {\bibinfo
  {volume} {90}},\ \bibinfo {pages} {100401} (\bibinfo {year}
  {2003})}\BibitemShut {NoStop}%
\bibitem [{\citenamefont {Altman}\ \emph {et~al.}(2003)\citenamefont {Altman},
  \citenamefont {Hofstetter}, \citenamefont {Demler},\ and\ \citenamefont
  {Lukin}}]{altman_phase_2003}%
  \BibitemOpen
  \bibfield  {author} {\bibinfo {author} {\bibfnamefont {E.}~\bibnamefont
  {Altman}}, \bibinfo {author} {\bibfnamefont {W.}~\bibnamefont {Hofstetter}},
  \bibinfo {author} {\bibfnamefont {E.}~\bibnamefont {Demler}}, \ and\ \bibinfo
  {author} {\bibfnamefont {M.~D.}\ \bibnamefont {Lukin}},\ }\href@noop {}
  {\bibfield  {journal} {\bibinfo  {journal} {New J. Phys.}\ }\textbf {\bibinfo
  {volume} {5}},\ \bibinfo {pages} {1} (\bibinfo {year} {2003})}\BibitemShut
  {NoStop}%
\bibitem [{\citenamefont {Stamper-Kurn}\ and\ \citenamefont
  {Ueda}(2013)}]{Ueda2013}%
  \BibitemOpen
  \bibfield  {author} {\bibinfo {author} {\bibfnamefont {D.~M.}\ \bibnamefont
  {Stamper-Kurn}}\ and\ \bibinfo {author} {\bibfnamefont {M.}~\bibnamefont
  {Ueda}},\ }\href {\doibase 10.1103/RevModPhys.85.1191} {\bibfield  {journal}
  {\bibinfo  {journal} {Rev. Mod. Phys.}\ }\textbf {\bibinfo {volume} {85}},\
  \bibinfo {pages} {1191} (\bibinfo {year} {2013})}\BibitemShut {NoStop}%
\bibitem [{\citenamefont {Hu}\ \emph {et~al.}(2016)\citenamefont {Hu},
  \citenamefont {Van~de Graaff}, \citenamefont {Kedar}, \citenamefont {Corson},
  \citenamefont {Cornell},\ and\ \citenamefont {Jin}}]{hu_bose_2016}%
  \BibitemOpen
  \bibfield  {author} {\bibinfo {author} {\bibfnamefont {M.-G.}\ \bibnamefont
  {Hu}}, \bibinfo {author} {\bibfnamefont {M.~J.}\ \bibnamefont {Van~de
  Graaff}}, \bibinfo {author} {\bibfnamefont {D.}~\bibnamefont {Kedar}},
  \bibinfo {author} {\bibfnamefont {J.~P.}\ \bibnamefont {Corson}}, \bibinfo
  {author} {\bibfnamefont {E.~A.}\ \bibnamefont {Cornell}}, \ and\ \bibinfo
  {author} {\bibfnamefont {D.~S.}\ \bibnamefont {Jin}},\ }\href {\doibase
  10.1103/PhysRevLett.117.055301} {\bibfield  {journal} {\bibinfo  {journal}
  {Physical Review Letters}\ }\textbf {\bibinfo {volume} {117}},\ \bibinfo
  {pages} {055301} (\bibinfo {year} {2016})}\BibitemShut {NoStop}%
\bibitem [{\citenamefont {J\o{}rgensen}\ \emph {et~al.}(2016)\citenamefont
  {J\o{}rgensen}, \citenamefont {Wacker}, \citenamefont {Skalmstang},
  \citenamefont {Parish}, \citenamefont {Levinsen}, \citenamefont
  {Christensen}, \citenamefont {Bruun},\ and\ \citenamefont
  {Arlt}}]{jorgensen_observation_2016}%
  \BibitemOpen
  \bibfield  {author} {\bibinfo {author} {\bibfnamefont {N.~B.}\ \bibnamefont
  {J\o{}rgensen}}, \bibinfo {author} {\bibfnamefont {L.}~\bibnamefont
  {Wacker}}, \bibinfo {author} {\bibfnamefont {K.~T.}\ \bibnamefont
  {Skalmstang}}, \bibinfo {author} {\bibfnamefont {M.~M.}\ \bibnamefont
  {Parish}}, \bibinfo {author} {\bibfnamefont {J.}~\bibnamefont {Levinsen}},
  \bibinfo {author} {\bibfnamefont {R.~S.}\ \bibnamefont {Christensen}},
  \bibinfo {author} {\bibfnamefont {G.~M.}\ \bibnamefont {Bruun}}, \ and\
  \bibinfo {author} {\bibfnamefont {J.~J.}\ \bibnamefont {Arlt}},\ }\href
  {\doibase 10.1103/PhysRevLett.117.055302} {\bibfield  {journal} {\bibinfo
  {journal} {Phys. Rev. Lett.}\ }\textbf {\bibinfo {volume} {117}},\ \bibinfo
  {pages} {055302} (\bibinfo {year} {2016})}\BibitemShut {NoStop}%
\bibitem [{\citenamefont {Cabrera}\ \emph {et~al.}(2018)\citenamefont
  {Cabrera}, \citenamefont {Tanzi}, \citenamefont {Sanz}, \citenamefont
  {Naylor}, \citenamefont {Thomas}, \citenamefont {Cheiney},\ and\
  \citenamefont {Tarruell}}]{cabrera_quantum_2018}%
  \BibitemOpen
  \bibfield  {author} {\bibinfo {author} {\bibfnamefont {C.~R.}\ \bibnamefont
  {Cabrera}}, \bibinfo {author} {\bibfnamefont {L.}~\bibnamefont {Tanzi}},
  \bibinfo {author} {\bibfnamefont {J.}~\bibnamefont {Sanz}}, \bibinfo {author}
  {\bibfnamefont {B.}~\bibnamefont {Naylor}}, \bibinfo {author} {\bibfnamefont
  {P.}~\bibnamefont {Thomas}}, \bibinfo {author} {\bibfnamefont
  {P.}~\bibnamefont {Cheiney}}, \ and\ \bibinfo {author} {\bibfnamefont
  {L.}~\bibnamefont {Tarruell}},\ }\href {\doibase 10.1126/science.aao5686}
  {\bibfield  {journal} {\bibinfo  {journal} {Science}\ }\textbf {\bibinfo
  {volume} {359}},\ \bibinfo {pages} {301} (\bibinfo {year}
  {2018})}\BibitemShut {NoStop}%
\bibitem [{\citenamefont {Semeghini}\ \emph {et~al.}(2018)\citenamefont
  {Semeghini}, \citenamefont {Ferioli}, \citenamefont {Masi}, \citenamefont
  {Mazzinghi}, \citenamefont {Wolswijk}, \citenamefont {Minardi}, \citenamefont
  {Modugno}, \citenamefont {Modugno}, \citenamefont {Inguscio},\ and\
  \citenamefont {Fattori}}]{Semeghini}%
  \BibitemOpen
  \bibfield  {author} {\bibinfo {author} {\bibfnamefont {G.}~\bibnamefont
  {Semeghini}}, \bibinfo {author} {\bibfnamefont {G.}~\bibnamefont {Ferioli}},
  \bibinfo {author} {\bibfnamefont {L.}~\bibnamefont {Masi}}, \bibinfo {author}
  {\bibfnamefont {C.}~\bibnamefont {Mazzinghi}}, \bibinfo {author}
  {\bibfnamefont {L.}~\bibnamefont {Wolswijk}}, \bibinfo {author}
  {\bibfnamefont {F.}~\bibnamefont {Minardi}}, \bibinfo {author} {\bibfnamefont
  {M.}~\bibnamefont {Modugno}}, \bibinfo {author} {\bibfnamefont
  {G.}~\bibnamefont {Modugno}}, \bibinfo {author} {\bibfnamefont
  {M.}~\bibnamefont {Inguscio}}, \ and\ \bibinfo {author} {\bibfnamefont
  {M.}~\bibnamefont {Fattori}},\ }\href {\doibase
  10.1103/PhysRevLett.120.235301} {\bibfield  {journal} {\bibinfo  {journal}
  {Phys. Rev. Lett.}\ }\textbf {\bibinfo {volume} {120}},\ \bibinfo {pages}
  {235301} (\bibinfo {year} {2018})}\BibitemShut {NoStop}%
\bibitem [{\citenamefont {Fava}\ \emph {et~al.}(2018)\citenamefont {Fava},
  \citenamefont {Bienaim\'e}, \citenamefont {Mordini}, \citenamefont {Colzi},
  \citenamefont {Qu}, \citenamefont {Stringari}, \citenamefont {Lamporesi},\
  and\ \citenamefont {Ferrari}}]{Fava2018}%
  \BibitemOpen
  \bibfield  {author} {\bibinfo {author} {\bibfnamefont {E.}~\bibnamefont
  {Fava}}, \bibinfo {author} {\bibfnamefont {T.}~\bibnamefont {Bienaim\'e}},
  \bibinfo {author} {\bibfnamefont {C.}~\bibnamefont {Mordini}}, \bibinfo
  {author} {\bibfnamefont {G.}~\bibnamefont {Colzi}}, \bibinfo {author}
  {\bibfnamefont {C.}~\bibnamefont {Qu}}, \bibinfo {author} {\bibfnamefont
  {S.}~\bibnamefont {Stringari}}, \bibinfo {author} {\bibfnamefont
  {G.}~\bibnamefont {Lamporesi}}, \ and\ \bibinfo {author} {\bibfnamefont
  {G.}~\bibnamefont {Ferrari}},\ }\href {\doibase
  10.1103/PhysRevLett.120.170401} {\bibfield  {journal} {\bibinfo  {journal}
  {Phys. Rev. Lett.}\ }\textbf {\bibinfo {volume} {120}},\ \bibinfo {pages}
  {170401} (\bibinfo {year} {2018})}\BibitemShut {NoStop}%
\bibitem [{\citenamefont {Beattie}\ \emph {et~al.}(2013)\citenamefont
  {Beattie}, \citenamefont {Moulder}, \citenamefont {Fletcher},\ and\
  \citenamefont {Hadzibabic}}]{Hadzibabic2013}%
  \BibitemOpen
  \bibfield  {author} {\bibinfo {author} {\bibfnamefont {S.}~\bibnamefont
  {Beattie}}, \bibinfo {author} {\bibfnamefont {S.}~\bibnamefont {Moulder}},
  \bibinfo {author} {\bibfnamefont {R.~J.}\ \bibnamefont {Fletcher}}, \ and\
  \bibinfo {author} {\bibfnamefont {Z.}~\bibnamefont {Hadzibabic}},\ }\href
  {\doibase 10.1103/PhysRevLett.110.025301} {\bibfield  {journal} {\bibinfo
  {journal} {Phys. Rev. Lett.}\ }\textbf {\bibinfo {volume} {110}},\ \bibinfo
  {pages} {025301} (\bibinfo {year} {2013})}\BibitemShut {NoStop}%
\bibitem [{\citenamefont {Abad}\ \emph {et~al.}(2014)\citenamefont {Abad},
  \citenamefont {Sartori}, \citenamefont {Finazzi},\ and\ \citenamefont
  {Recati}}]{Abad2014}%
  \BibitemOpen
  \bibfield  {author} {\bibinfo {author} {\bibfnamefont {M.}~\bibnamefont
  {Abad}}, \bibinfo {author} {\bibfnamefont {A.}~\bibnamefont {Sartori}},
  \bibinfo {author} {\bibfnamefont {S.}~\bibnamefont {Finazzi}}, \ and\
  \bibinfo {author} {\bibfnamefont {A.}~\bibnamefont {Recati}},\ }\href
  {\doibase 10.1103/PhysRevA.89.053602} {\bibfield  {journal} {\bibinfo
  {journal} {Phys. Rev. A}\ }\textbf {\bibinfo {volume} {89}},\ \bibinfo
  {pages} {053602} (\bibinfo {year} {2014})}\BibitemShut {NoStop}%
\bibitem [{\citenamefont {Ni}\ \emph {et~al.}(2008)\citenamefont {Ni},
  \citenamefont {Ospelkaus}, \citenamefont {de~Miranda}, \citenamefont
  {Pe{\textquoteright}er}, \citenamefont {Neyenhuis}, \citenamefont {Zirbel},
  \citenamefont {Kotochigova}, \citenamefont {Julienne}, \citenamefont {Jin},\
  and\ \citenamefont {Ye}}]{Jin2008}%
  \BibitemOpen
  \bibfield  {author} {\bibinfo {author} {\bibfnamefont {K.-K.}\ \bibnamefont
  {Ni}}, \bibinfo {author} {\bibfnamefont {S.}~\bibnamefont {Ospelkaus}},
  \bibinfo {author} {\bibfnamefont {M.~H.~G.}\ \bibnamefont {de~Miranda}},
  \bibinfo {author} {\bibfnamefont {A.}~\bibnamefont {Pe{\textquoteright}er}},
  \bibinfo {author} {\bibfnamefont {B.}~\bibnamefont {Neyenhuis}}, \bibinfo
  {author} {\bibfnamefont {J.~J.}\ \bibnamefont {Zirbel}}, \bibinfo {author}
  {\bibfnamefont {S.}~\bibnamefont {Kotochigova}}, \bibinfo {author}
  {\bibfnamefont {P.~S.}\ \bibnamefont {Julienne}}, \bibinfo {author}
  {\bibfnamefont {D.~S.}\ \bibnamefont {Jin}}, \ and\ \bibinfo {author}
  {\bibfnamefont {J.}~\bibnamefont {Ye}},\ }\href {\doibase
  10.1126/science.1163861} {\bibfield  {journal} {\bibinfo  {journal}
  {Science}\ }\textbf {\bibinfo {volume} {322}},\ \bibinfo {pages} {231}
  (\bibinfo {year} {2008})}\BibitemShut {NoStop}%
\bibitem [{\citenamefont {Molony}\ \emph {et~al.}(2014)\citenamefont {Molony},
  \citenamefont {Gregory}, \citenamefont {Ji}, \citenamefont {Lu},
  \citenamefont {K\"oppinger}, \citenamefont {Le~Sueur}, \citenamefont
  {Blackley}, \citenamefont {Hutson},\ and\ \citenamefont
  {Cornish}}]{Cornish2014}%
  \BibitemOpen
  \bibfield  {author} {\bibinfo {author} {\bibfnamefont {P.~K.}\ \bibnamefont
  {Molony}}, \bibinfo {author} {\bibfnamefont {P.~D.}\ \bibnamefont {Gregory}},
  \bibinfo {author} {\bibfnamefont {Z.}~\bibnamefont {Ji}}, \bibinfo {author}
  {\bibfnamefont {B.}~\bibnamefont {Lu}}, \bibinfo {author} {\bibfnamefont
  {M.~P.}\ \bibnamefont {K\"oppinger}}, \bibinfo {author} {\bibfnamefont
  {C.~R.}\ \bibnamefont {Le~Sueur}}, \bibinfo {author} {\bibfnamefont {C.~L.}\
  \bibnamefont {Blackley}}, \bibinfo {author} {\bibfnamefont {J.~M.}\
  \bibnamefont {Hutson}}, \ and\ \bibinfo {author} {\bibfnamefont {S.~L.}\
  \bibnamefont {Cornish}},\ }\href {\doibase 10.1103/PhysRevLett.113.255301}
  {\bibfield  {journal} {\bibinfo  {journal} {Phys. Rev. Lett.}\ }\textbf
  {\bibinfo {volume} {113}},\ \bibinfo {pages} {255301} (\bibinfo {year}
  {2014})}\BibitemShut {NoStop}%
\bibitem [{\citenamefont {Takekoshi}\ \emph {et~al.}(2014)\citenamefont
  {Takekoshi}, \citenamefont {Reichs\"ollner}, \citenamefont {Schindewolf},
  \citenamefont {Hutson}, \citenamefont {Le~Sueur}, \citenamefont {Dulieu},
  \citenamefont {Ferlaino}, \citenamefont {Grimm},\ and\ \citenamefont
  {N\"agerl}}]{Nagerl2014}%
  \BibitemOpen
  \bibfield  {author} {\bibinfo {author} {\bibfnamefont {T.}~\bibnamefont
  {Takekoshi}}, \bibinfo {author} {\bibfnamefont {L.}~\bibnamefont
  {Reichs\"ollner}}, \bibinfo {author} {\bibfnamefont {A.}~\bibnamefont
  {Schindewolf}}, \bibinfo {author} {\bibfnamefont {J.~M.}\ \bibnamefont
  {Hutson}}, \bibinfo {author} {\bibfnamefont {C.~R.}\ \bibnamefont
  {Le~Sueur}}, \bibinfo {author} {\bibfnamefont {O.}~\bibnamefont {Dulieu}},
  \bibinfo {author} {\bibfnamefont {F.}~\bibnamefont {Ferlaino}}, \bibinfo
  {author} {\bibfnamefont {R.}~\bibnamefont {Grimm}}, \ and\ \bibinfo {author}
  {\bibfnamefont {H.-C.}\ \bibnamefont {N\"agerl}},\ }\href {\doibase
  10.1103/PhysRevLett.113.205301} {\bibfield  {journal} {\bibinfo  {journal}
  {Phys. Rev. Lett.}\ }\textbf {\bibinfo {volume} {113}},\ \bibinfo {pages}
  {205301} (\bibinfo {year} {2014})}\BibitemShut {NoStop}%
\bibitem [{\citenamefont {Guo}\ \emph {et~al.}(2016)\citenamefont {Guo},
  \citenamefont {Zhu}, \citenamefont {Lu}, \citenamefont {Ye}, \citenamefont
  {Wang}, \citenamefont {Vexiau}, \citenamefont {Bouloufa-Maafa}, \citenamefont
  {Qu\'em\'ener}, \citenamefont {Dulieu},\ and\ \citenamefont
  {Wang}}]{Dulieu2016}%
  \BibitemOpen
  \bibfield  {author} {\bibinfo {author} {\bibfnamefont {M.}~\bibnamefont
  {Guo}}, \bibinfo {author} {\bibfnamefont {B.}~\bibnamefont {Zhu}}, \bibinfo
  {author} {\bibfnamefont {B.}~\bibnamefont {Lu}}, \bibinfo {author}
  {\bibfnamefont {X.}~\bibnamefont {Ye}}, \bibinfo {author} {\bibfnamefont
  {F.}~\bibnamefont {Wang}}, \bibinfo {author} {\bibfnamefont {R.}~\bibnamefont
  {Vexiau}}, \bibinfo {author} {\bibfnamefont {N.}~\bibnamefont
  {Bouloufa-Maafa}}, \bibinfo {author} {\bibfnamefont {G.}~\bibnamefont
  {Qu\'em\'ener}}, \bibinfo {author} {\bibfnamefont {O.}~\bibnamefont
  {Dulieu}}, \ and\ \bibinfo {author} {\bibfnamefont {D.}~\bibnamefont
  {Wang}},\ }\href {\doibase 10.1103/PhysRevLett.116.205303} {\bibfield
  {journal} {\bibinfo  {journal} {Phys. Rev. Lett.}\ }\textbf {\bibinfo
  {volume} {116}},\ \bibinfo {pages} {205303} (\bibinfo {year}
  {2016})}\BibitemShut {NoStop}%
\bibitem [{\citenamefont {{De Marco}}\ \emph {et~al.}(2018)\citenamefont {{De
  Marco}}, \citenamefont {{Valtolina}}, \citenamefont {{Matsuda}},
  \citenamefont {{Tobias}}, \citenamefont {{Covey}},\ and\ \citenamefont
  {{Ye}}}]{JunYe2018}%
  \BibitemOpen
  \bibfield  {author} {\bibinfo {author} {\bibfnamefont {L.}~\bibnamefont {{De
  Marco}}}, \bibinfo {author} {\bibfnamefont {G.}~\bibnamefont {{Valtolina}}},
  \bibinfo {author} {\bibfnamefont {K.}~\bibnamefont {{Matsuda}}}, \bibinfo
  {author} {\bibfnamefont {W.~G.}\ \bibnamefont {{Tobias}}}, \bibinfo {author}
  {\bibfnamefont {J.~P.}\ \bibnamefont {{Covey}}}, \ and\ \bibinfo {author}
  {\bibfnamefont {J.}~\bibnamefont {{Ye}}},\ }\href@noop {} {\bibfield
  {journal} {\bibinfo  {journal} {ArXiv e-prints}\ } (\bibinfo {year}
  {2018})},\ \Eprint {http://arxiv.org/abs/1808.00028} {arXiv:1808.00028
  [physics.atom-ph]} \BibitemShut {NoStop}%
\bibitem [{\citenamefont {Modugno}\ \emph {et~al.}(2001)\citenamefont
  {Modugno}, \citenamefont {Ferrari}, \citenamefont {Roati}, \citenamefont
  {Brecha}, \citenamefont {Simoni},\ and\ \citenamefont
  {Inguscio}}]{Modugno2001}%
  \BibitemOpen
  \bibfield  {author} {\bibinfo {author} {\bibfnamefont {G.}~\bibnamefont
  {Modugno}}, \bibinfo {author} {\bibfnamefont {G.}~\bibnamefont {Ferrari}},
  \bibinfo {author} {\bibfnamefont {G.}~\bibnamefont {Roati}}, \bibinfo
  {author} {\bibfnamefont {R.~J.}\ \bibnamefont {Brecha}}, \bibinfo {author}
  {\bibfnamefont {A.}~\bibnamefont {Simoni}}, \ and\ \bibinfo {author}
  {\bibfnamefont {M.}~\bibnamefont {Inguscio}},\ }\href {\doibase
  10.1126/science.1066687} {\bibfield  {journal} {\bibinfo  {journal}
  {Science}\ }\textbf {\bibinfo {volume} {294}},\ \bibinfo {pages} {1320}
  (\bibinfo {year} {2001})}\BibitemShut {NoStop}%
\bibitem [{\citenamefont {Lin}\ \emph {et~al.}(2009)\citenamefont {Lin},
  \citenamefont {Perry}, \citenamefont {Compton}, \citenamefont {Spielman},\
  and\ \citenamefont {Porto}}]{lin_rapid_2009}%
  \BibitemOpen
  \bibfield  {author} {\bibinfo {author} {\bibfnamefont {Y.-J.}\ \bibnamefont
  {Lin}}, \bibinfo {author} {\bibfnamefont {A.~R.}\ \bibnamefont {Perry}},
  \bibinfo {author} {\bibfnamefont {R.~L.}\ \bibnamefont {Compton}}, \bibinfo
  {author} {\bibfnamefont {I.~B.}\ \bibnamefont {Spielman}}, \ and\ \bibinfo
  {author} {\bibfnamefont {J.~V.}\ \bibnamefont {Porto}},\ }\href {\doibase
  10.1103/PhysRevA.79.063631} {\bibfield  {journal} {\bibinfo  {journal}
  {Physical Review A}\ }\textbf {\bibinfo {volume} {79}},\ \bibinfo {pages}
  {063631} (\bibinfo {year} {2009})}\BibitemShut {NoStop}%
\bibitem [{\citenamefont {Barnett}\ \emph {et~al.}(2008)\citenamefont
  {Barnett}, \citenamefont {Refael}, \citenamefont {Porter},\ and\
  \citenamefont {B\"uchler}}]{Barnett2008}%
  \BibitemOpen
  \bibfield  {author} {\bibinfo {author} {\bibfnamefont {R.}~\bibnamefont
  {Barnett}}, \bibinfo {author} {\bibfnamefont {G.}~\bibnamefont {Refael}},
  \bibinfo {author} {\bibfnamefont {M.~A.}\ \bibnamefont {Porter}}, \ and\
  \bibinfo {author} {\bibfnamefont {H.~P.}\ \bibnamefont {B\"uchler}},\ }\href
  {http://stacks.iop.org/1367-2630/10/i=4/a=043030} {\bibfield  {journal}
  {\bibinfo  {journal} {New J. Phys.}\ }\textbf {\bibinfo {volume} {10}},\
  \bibinfo {pages} {043030} (\bibinfo {year} {2008})}\BibitemShut {NoStop}%
\bibitem [{\citenamefont {Kuopanportti}\ \emph {et~al.}(2012)\citenamefont
  {Kuopanportti}, \citenamefont {Huhtam\"aki},\ and\ \citenamefont
  {M\"ott\"onen}}]{Kuopanportti2012}%
  \BibitemOpen
  \bibfield  {author} {\bibinfo {author} {\bibfnamefont {P.}~\bibnamefont
  {Kuopanportti}}, \bibinfo {author} {\bibfnamefont {J.~A.~M.}\ \bibnamefont
  {Huhtam\"aki}}, \ and\ \bibinfo {author} {\bibfnamefont {M.}~\bibnamefont
  {M\"ott\"onen}},\ }\href {\doibase 10.1103/PhysRevA.85.043613} {\bibfield
  {journal} {\bibinfo  {journal} {Phys. Rev. A}\ }\textbf {\bibinfo {volume}
  {85}},\ \bibinfo {pages} {043613} (\bibinfo {year} {2012})}\BibitemShut
  {NoStop}%
\bibitem [{\citenamefont {Kuopanportti}\ \emph {et~al.}(2015)\citenamefont
  {Kuopanportti}, \citenamefont {Orlova},\ and\ \citenamefont {Milo\ifmmode
  \check{s}\else \v{s}\fi{}evi\ifmmode~\acute{c}\else
  \'{c}\fi{}}}]{Kuopanportti2015}%
  \BibitemOpen
  \bibfield  {author} {\bibinfo {author} {\bibfnamefont {P.}~\bibnamefont
  {Kuopanportti}}, \bibinfo {author} {\bibfnamefont {N.~V.}\ \bibnamefont
  {Orlova}}, \ and\ \bibinfo {author} {\bibfnamefont {M.~V.}\ \bibnamefont
  {Milo\ifmmode \check{s}\else \v{s}\fi{}evi\ifmmode~\acute{c}\else
  \'{c}\fi{}}},\ }\href {\doibase 10.1103/PhysRevA.91.043605} {\bibfield
  {journal} {\bibinfo  {journal} {Phys. Rev. A}\ }\textbf {\bibinfo {volume}
  {91}},\ \bibinfo {pages} {043605} (\bibinfo {year} {2015})}\BibitemShut
  {NoStop}%
\bibitem [{\citenamefont {Petrov}(2015)}]{Petrov2015}%
  \BibitemOpen
  \bibfield  {author} {\bibinfo {author} {\bibfnamefont {D.~S.}\ \bibnamefont
  {Petrov}},\ }\href {\doibase 10.1103/PhysRevLett.115.155302} {\bibfield
  {journal} {\bibinfo  {journal} {Phys. Rev. Lett.}\ }\textbf {\bibinfo
  {volume} {115}},\ \bibinfo {pages} {155302} (\bibinfo {year}
  {2015})}\BibitemShut {NoStop}%
\bibitem [{\citenamefont {Fil}\ and\ \citenamefont
  {Shevchenko}(2005)}]{Shevchenko2005}%
  \BibitemOpen
  \bibfield  {author} {\bibinfo {author} {\bibfnamefont {D.~V.}\ \bibnamefont
  {Fil}}\ and\ \bibinfo {author} {\bibfnamefont {S.~I.}\ \bibnamefont
  {Shevchenko}},\ }\href {\doibase 10.1103/PhysRevA.72.013616} {\bibfield
  {journal} {\bibinfo  {journal} {Phys. Rev. A}\ }\textbf {\bibinfo {volume}
  {72}},\ \bibinfo {pages} {013616} (\bibinfo {year} {2005})}\BibitemShut
  {NoStop}%
\bibitem [{\citenamefont {Nespolo}\ \emph {et~al.}(2017)\citenamefont
  {Nespolo}, \citenamefont {Astrakharchik},\ and\ \citenamefont
  {Recati}}]{Nespolo2017}%
  \BibitemOpen
  \bibfield  {author} {\bibinfo {author} {\bibfnamefont {J.}~\bibnamefont
  {Nespolo}}, \bibinfo {author} {\bibfnamefont {G.~E.}\ \bibnamefont
  {Astrakharchik}}, \ and\ \bibinfo {author} {\bibfnamefont {A.}~\bibnamefont
  {Recati}},\ }\href {http://stacks.iop.org/1367-2630/19/i=12/a=125005}
  {\bibfield  {journal} {\bibinfo  {journal} {New J. Phys.}\ }\textbf {\bibinfo
  {volume} {19}},\ \bibinfo {pages} {125005} (\bibinfo {year}
  {2017})}\BibitemShut {NoStop}%
\bibitem [{\citenamefont {Sellin}\ and\ \citenamefont
  {Babaev}(2018)}]{Babaev2018}%
  \BibitemOpen
  \bibfield  {author} {\bibinfo {author} {\bibfnamefont {K.}~\bibnamefont
  {Sellin}}\ and\ \bibinfo {author} {\bibfnamefont {E.}~\bibnamefont
  {Babaev}},\ }\href {\doibase 10.1103/PhysRevB.97.094517} {\bibfield
  {journal} {\bibinfo  {journal} {Phys. Rev. B}\ }\textbf {\bibinfo {volume}
  {97}},\ \bibinfo {pages} {094517} (\bibinfo {year} {2018})}\BibitemShut
  {NoStop}%
\bibitem [{\citenamefont {Parisi}\ \emph {et~al.}(2018)\citenamefont {Parisi},
  \citenamefont {Astrakharchik},\ and\ \citenamefont
  {Giorgini}}]{Giorgini2018}%
  \BibitemOpen
  \bibfield  {author} {\bibinfo {author} {\bibfnamefont {L.}~\bibnamefont
  {Parisi}}, \bibinfo {author} {\bibfnamefont {G.~E.}\ \bibnamefont
  {Astrakharchik}}, \ and\ \bibinfo {author} {\bibfnamefont {S.}~\bibnamefont
  {Giorgini}},\ }\href {\doibase 10.1103/PhysRevLett.121.025302} {\bibfield
  {journal} {\bibinfo  {journal} {Phys. Rev. Lett.}\ }\textbf {\bibinfo
  {volume} {121}},\ \bibinfo {pages} {025302} (\bibinfo {year}
  {2018})}\BibitemShut {NoStop}%
\bibitem [{\citenamefont {Kleine~B\"uning}\ \emph {et~al.}(2011)\citenamefont
  {Kleine~B\"uning}, \citenamefont {Will}, \citenamefont {Ertmer},
  \citenamefont {Rasel}, \citenamefont {Arlt}, \citenamefont {Klempt},
  \citenamefont {Ramirez-Martinez}, \citenamefont {Pi\'echon},\ and\
  \citenamefont {Rosenbusch}}]{PhysRevLett.106.240801}%
  \BibitemOpen
  \bibfield  {author} {\bibinfo {author} {\bibfnamefont {G.}~\bibnamefont
  {Kleine~B\"uning}}, \bibinfo {author} {\bibfnamefont {J.}~\bibnamefont
  {Will}}, \bibinfo {author} {\bibfnamefont {W.}~\bibnamefont {Ertmer}},
  \bibinfo {author} {\bibfnamefont {E.}~\bibnamefont {Rasel}}, \bibinfo
  {author} {\bibfnamefont {J.}~\bibnamefont {Arlt}}, \bibinfo {author}
  {\bibfnamefont {C.}~\bibnamefont {Klempt}}, \bibinfo {author} {\bibfnamefont
  {F.}~\bibnamefont {Ramirez-Martinez}}, \bibinfo {author} {\bibfnamefont
  {F.}~\bibnamefont {Pi\'echon}}, \ and\ \bibinfo {author} {\bibfnamefont
  {P.}~\bibnamefont {Rosenbusch}},\ }\href {\doibase
  10.1103/PhysRevLett.106.240801} {\bibfield  {journal} {\bibinfo  {journal}
  {Phys. Rev. Lett.}\ }\textbf {\bibinfo {volume} {106}},\ \bibinfo {pages}
  {240801} (\bibinfo {year} {2011})}\BibitemShut {NoStop}%
\bibitem [{\citenamefont {De~Sarlo}\ \emph {et~al.}(2007)\citenamefont
  {De~Sarlo}, \citenamefont {Maioli}, \citenamefont {Barontini}, \citenamefont
  {Catani}, \citenamefont {Minardi},\ and\ \citenamefont
  {Inguscio}}]{de_sarlo_collisional_2007}%
  \BibitemOpen
  \bibfield  {author} {\bibinfo {author} {\bibfnamefont {L.}~\bibnamefont
  {De~Sarlo}}, \bibinfo {author} {\bibfnamefont {P.}~\bibnamefont {Maioli}},
  \bibinfo {author} {\bibfnamefont {G.}~\bibnamefont {Barontini}}, \bibinfo
  {author} {\bibfnamefont {J.}~\bibnamefont {Catani}}, \bibinfo {author}
  {\bibfnamefont {F.}~\bibnamefont {Minardi}}, \ and\ \bibinfo {author}
  {\bibfnamefont {M.}~\bibnamefont {Inguscio}},\ }\href {\doibase
  10.1103/PhysRevA.75.022715} {\bibfield  {journal} {\bibinfo  {journal} {Phys.
  Rev. A}\ }\textbf {\bibinfo {volume} {75}},\ \bibinfo {pages} {022715}
  (\bibinfo {year} {2007})}\BibitemShut {NoStop}%
\bibitem [{\citenamefont {Campbell}\ \emph {et~al.}(2010)\citenamefont
  {Campbell}, \citenamefont {Smith}, \citenamefont {Tammuz}, \citenamefont
  {Beattie}, \citenamefont {Moulder},\ and\ \citenamefont
  {Hadzibabic}}]{campbell_efficient_2010}%
  \BibitemOpen
  \bibfield  {author} {\bibinfo {author} {\bibfnamefont {R.~L.~D.}\
  \bibnamefont {Campbell}}, \bibinfo {author} {\bibfnamefont {R.~P.}\
  \bibnamefont {Smith}}, \bibinfo {author} {\bibfnamefont {N.}~\bibnamefont
  {Tammuz}}, \bibinfo {author} {\bibfnamefont {S.}~\bibnamefont {Beattie}},
  \bibinfo {author} {\bibfnamefont {S.}~\bibnamefont {Moulder}}, \ and\
  \bibinfo {author} {\bibfnamefont {Z.}~\bibnamefont {Hadzibabic}},\ }\href
  {\doibase 10.1103/PhysRevA.82.063611} {\bibfield  {journal} {\bibinfo
  {journal} {Phys. Rev. A}\ }\textbf {\bibinfo {volume} {82}},\ \bibinfo
  {pages} {063611} (\bibinfo {year} {2010})}\BibitemShut {NoStop}%
\bibitem [{\citenamefont {Catani}\ \emph {et~al.}(2008)\citenamefont {Catani},
  \citenamefont {De~Sarlo}, \citenamefont {Barontini}, \citenamefont
  {Minardi},\ and\ \citenamefont {Inguscio}}]{Catani2008}%
  \BibitemOpen
  \bibfield  {author} {\bibinfo {author} {\bibfnamefont {J.}~\bibnamefont
  {Catani}}, \bibinfo {author} {\bibfnamefont {L.}~\bibnamefont {De~Sarlo}},
  \bibinfo {author} {\bibfnamefont {G.}~\bibnamefont {Barontini}}, \bibinfo
  {author} {\bibfnamefont {F.}~\bibnamefont {Minardi}}, \ and\ \bibinfo
  {author} {\bibfnamefont {M.}~\bibnamefont {Inguscio}},\ }\href {\doibase
  10.1103/PhysRevA.77.011603} {\bibfield  {journal} {\bibinfo  {journal} {Phys.
  Rev. A}\ }\textbf {\bibinfo {volume} {77}},\ \bibinfo {pages} {011603}
  (\bibinfo {year} {2008})}\BibitemShut {NoStop}%
\bibitem [{\citenamefont {Petrich}\ \emph {et~al.}(1995)\citenamefont
  {Petrich}, \citenamefont {Anderson}, \citenamefont {Ensher},\ and\
  \citenamefont {Cornell}}]{Petrich1995}%
  \BibitemOpen
  \bibfield  {author} {\bibinfo {author} {\bibfnamefont {W.}~\bibnamefont
  {Petrich}}, \bibinfo {author} {\bibfnamefont {M.~H.}\ \bibnamefont
  {Anderson}}, \bibinfo {author} {\bibfnamefont {J.~R.}\ \bibnamefont
  {Ensher}}, \ and\ \bibinfo {author} {\bibfnamefont {E.~A.}\ \bibnamefont
  {Cornell}},\ }\href {\doibase 10.1103/PhysRevLett.74.3352} {\bibfield
  {journal} {\bibinfo  {journal} {Phys. Rev. Lett.}\ }\textbf {\bibinfo
  {volume} {74}},\ \bibinfo {pages} {3352} (\bibinfo {year}
  {1995})}\BibitemShut {NoStop}%
\bibitem [{\citenamefont {Dubessy}\ \emph {et~al.}(2012)\citenamefont
  {Dubessy}, \citenamefont {Merloti}, \citenamefont {Longchambon},
  \citenamefont {Pottie}, \citenamefont {Liennard}, \citenamefont {Perrin},
  \citenamefont {Lorent},\ and\ \citenamefont {Perrin}}]{Dubessy2012}%
  \BibitemOpen
  \bibfield  {author} {\bibinfo {author} {\bibfnamefont {R.}~\bibnamefont
  {Dubessy}}, \bibinfo {author} {\bibfnamefont {K.}~\bibnamefont {Merloti}},
  \bibinfo {author} {\bibfnamefont {L.}~\bibnamefont {Longchambon}}, \bibinfo
  {author} {\bibfnamefont {P.-E.}\ \bibnamefont {Pottie}}, \bibinfo {author}
  {\bibfnamefont {T.}~\bibnamefont {Liennard}}, \bibinfo {author}
  {\bibfnamefont {A.}~\bibnamefont {Perrin}}, \bibinfo {author} {\bibfnamefont
  {V.}~\bibnamefont {Lorent}}, \ and\ \bibinfo {author} {\bibfnamefont
  {H.}~\bibnamefont {Perrin}},\ }\href {\doibase 10.1103/PhysRevA.85.013643}
  {\bibfield  {journal} {\bibinfo  {journal} {Phys. Rev. A}\ }\textbf {\bibinfo
  {volume} {85}},\ \bibinfo {pages} {013643} (\bibinfo {year}
  {2012})}\BibitemShut {NoStop}%
\bibitem [{\citenamefont {Heo}\ \emph {et~al.}(2011)\citenamefont {Heo},
  \citenamefont {Choi},\ and\ \citenamefont {Shin}}]{Heo2011}%
  \BibitemOpen
  \bibfield  {author} {\bibinfo {author} {\bibfnamefont {M.~S.}\ \bibnamefont
  {Heo}}, \bibinfo {author} {\bibfnamefont {J.~Y.}\ \bibnamefont {Choi}}, \
  and\ \bibinfo {author} {\bibfnamefont {Y.~I.}\ \bibnamefont {Shin}},\ }\href
  {\doibase 10.1103/PhysRevA.83.013622} {\bibfield  {journal} {\bibinfo
  {journal} {Phys. Rev. A}\ }\textbf {\bibinfo {volume} {83}},\ \bibinfo
  {pages} {013622} (\bibinfo {year} {2011})}\BibitemShut {NoStop}%
\bibitem [{\citenamefont {Simoni}\ \emph {et~al.}(2008)\citenamefont {Simoni},
  \citenamefont {Zaccanti}, \citenamefont {D'Errico}, \citenamefont {Fattori},
  \citenamefont {Roati}, \citenamefont {Inguscio},\ and\ \citenamefont
  {Modugno}}]{simoni2008}%
  \BibitemOpen
  \bibfield  {author} {\bibinfo {author} {\bibfnamefont {A.}~\bibnamefont
  {Simoni}}, \bibinfo {author} {\bibfnamefont {M.}~\bibnamefont {Zaccanti}},
  \bibinfo {author} {\bibfnamefont {C.}~\bibnamefont {D'Errico}}, \bibinfo
  {author} {\bibfnamefont {M.}~\bibnamefont {Fattori}}, \bibinfo {author}
  {\bibfnamefont {G.}~\bibnamefont {Roati}}, \bibinfo {author} {\bibfnamefont
  {M.}~\bibnamefont {Inguscio}}, \ and\ \bibinfo {author} {\bibfnamefont
  {G.}~\bibnamefont {Modugno}},\ }\href {\doibase 10.1103/PhysRevA.77.052705}
  {\bibfield  {journal} {\bibinfo  {journal} {Phys. Rev. A}\ }\textbf {\bibinfo
  {volume} {77}},\ \bibinfo {pages} {052705} (\bibinfo {year}
  {2008})}\BibitemShut {NoStop}%
\bibitem [{Note1()}]{Note1}%
  \BibitemOpen
  \bibinfo {note} {We develop a $^{87}$Rb collision model based on the accurate
  potential parameters determined in \cite {vanKepen2002}. In addition to the
  well-known dipolar coupling, our model also includes the second-order
  spin-orbit interaction, first discussed in the context of ultracold gases in
  \cite {Krauss1996}. The atom-loss rate is found to be sensitive to the
  variation of collision energy and magnetic field, and in the range of our
  experimental parameters is predicted to vary between $1 \times 10^{-15}$ and
  $2 \times 10^{-14}$ cm$^3$/s.}\BibitemShut {Stop}%
\bibitem [{\citenamefont {Haas}\ \emph {et~al.}(2007)\citenamefont {Haas},
  \citenamefont {Leung}, \citenamefont {Frese}, \citenamefont {Haubrich},
  \citenamefont {John}, \citenamefont {Weber}, \citenamefont {Rauschenbeutel},\
  and\ \citenamefont {Meschede}}]{Meschede2007}%
  \BibitemOpen
  \bibfield  {author} {\bibinfo {author} {\bibfnamefont {M.}~\bibnamefont
  {Haas}}, \bibinfo {author} {\bibfnamefont {V.}~\bibnamefont {Leung}},
  \bibinfo {author} {\bibfnamefont {D.}~\bibnamefont {Frese}}, \bibinfo
  {author} {\bibfnamefont {D.}~\bibnamefont {Haubrich}}, \bibinfo {author}
  {\bibfnamefont {S.}~\bibnamefont {John}}, \bibinfo {author} {\bibfnamefont
  {C.}~\bibnamefont {Weber}}, \bibinfo {author} {\bibfnamefont
  {A.}~\bibnamefont {Rauschenbeutel}}, \ and\ \bibinfo {author} {\bibfnamefont
  {D.}~\bibnamefont {Meschede}},\ }\href
  {http://stacks.iop.org/1367-2630/9/i=5/a=147} {\bibfield  {journal} {\bibinfo
   {journal} {New Journal of Physics}\ }\textbf {\bibinfo {volume} {9}},\
  \bibinfo {pages} {147} (\bibinfo {year} {2007})}\BibitemShut {NoStop}%
\bibitem [{Note2()}]{Note2}%
  \BibitemOpen
  \bibinfo {note} {The atom number $N$ of both $^{41}$K and $^{87}$Rb have been
  calibrated using the saturation absorption imaging technique, described in
  \cite {Reinaudi:07}. We estimate an uncertainty in $N$ of 35\% for both
  atomic species.}\BibitemShut {Stop}%
\bibitem [{\citenamefont {Riboli}\ and\ \citenamefont
  {Modugno}(2002)}]{MicheleModugno2002}%
  \BibitemOpen
  \bibfield  {author} {\bibinfo {author} {\bibfnamefont {F.}~\bibnamefont
  {Riboli}}\ and\ \bibinfo {author} {\bibfnamefont {M.}~\bibnamefont
  {Modugno}},\ }\href {\doibase 10.1103/PhysRevA.65.063614} {\bibfield
  {journal} {\bibinfo  {journal} {Phys. Rev. A}\ }\textbf {\bibinfo {volume}
  {65}},\ \bibinfo {pages} {063614} (\bibinfo {year} {2002})}\BibitemShut
  {NoStop}%
\bibitem [{Note3()}]{Note3}%
  \BibitemOpen
  \bibinfo {note} {A set of two three-dimensional Gross-Pitaevskii (GP)
  equations is considered \cite {Pethick}. The ground state is found by using a
  standard imaginary time evolution \cite {DalfovoRev1999}. The time-dependent
  GP equations are solved by means of a split-step method that makes use of
  fast Fourier transforms \cite {Jackson1998}}\BibitemShut {NoStop}%
\bibitem [{\citenamefont {van Kempen}\ \emph {et~al.}(2002)\citenamefont {van
  Kempen}, \citenamefont {Kokkelmans}, \citenamefont {Heinzen},\ and\
  \citenamefont {Verhaar}}]{vanKepen2002}%
  \BibitemOpen
  \bibfield  {author} {\bibinfo {author} {\bibfnamefont {E.~G.~M.}\
  \bibnamefont {van Kempen}}, \bibinfo {author} {\bibfnamefont {S.~J. J.
  M.~F.}\ \bibnamefont {Kokkelmans}}, \bibinfo {author} {\bibfnamefont {D.~J.}\
  \bibnamefont {Heinzen}}, \ and\ \bibinfo {author} {\bibfnamefont {B.~J.}\
  \bibnamefont {Verhaar}},\ }\href {\doibase 10.1103/PhysRevLett.88.093201}
  {\bibfield  {journal} {\bibinfo  {journal} {Phys. Rev. Lett.}\ }\textbf
  {\bibinfo {volume} {88}},\ \bibinfo {pages} {093201} (\bibinfo {year}
  {2002})}\BibitemShut {NoStop}%
\bibitem [{\citenamefont {Mies}\ \emph {et~al.}(1996)\citenamefont {Mies},
  \citenamefont {Williams}, \citenamefont {Julienne},\ and\ \citenamefont
  {Krauss}}]{Krauss1996}%
  \BibitemOpen
  \bibfield  {author} {\bibinfo {author} {\bibfnamefont {F.~H.}\ \bibnamefont
  {Mies}}, \bibinfo {author} {\bibfnamefont {C.~J.}\ \bibnamefont {Williams}},
  \bibinfo {author} {\bibfnamefont {P.~S.}\ \bibnamefont {Julienne}}, \ and\
  \bibinfo {author} {\bibfnamefont {M.}~\bibnamefont {Krauss}},\ }\href
  {http://doi.org/10.6028/jres.101.052} {\bibfield  {journal} {\bibinfo
  {journal} {J. Res. Natl. Inst. Stand. Technol.}\ }\textbf {\bibinfo {volume}
  {101}},\ \bibinfo {pages} {521} (\bibinfo {year} {1996})}\BibitemShut
  {NoStop}%
\bibitem [{\citenamefont {Reinaudi}\ \emph {et~al.}(2007)\citenamefont
  {Reinaudi}, \citenamefont {Lahaye}, \citenamefont {Wang},\ and\ \citenamefont
  {Gu\'{e}ry-Odelin}}]{Reinaudi:07}%
  \BibitemOpen
  \bibfield  {author} {\bibinfo {author} {\bibfnamefont {G.}~\bibnamefont
  {Reinaudi}}, \bibinfo {author} {\bibfnamefont {T.}~\bibnamefont {Lahaye}},
  \bibinfo {author} {\bibfnamefont {Z.}~\bibnamefont {Wang}}, \ and\ \bibinfo
  {author} {\bibfnamefont {D.}~\bibnamefont {Gu\'{e}ry-Odelin}},\ }\href
  {\doibase 10.1364/OL.32.003143} {\bibfield  {journal} {\bibinfo  {journal}
  {Opt. Lett.}\ }\textbf {\bibinfo {volume} {32}},\ \bibinfo {pages} {3143}
  (\bibinfo {year} {2007})}\BibitemShut {NoStop}%
\bibitem [{\citenamefont {Pethick}\ and\ \citenamefont
  {Smith}(2008)}]{Pethick}%
  \BibitemOpen
  \bibfield  {author} {\bibinfo {author} {\bibfnamefont {C.~J.}\ \bibnamefont
  {Pethick}}\ and\ \bibinfo {author} {\bibfnamefont {H.}~\bibnamefont
  {Smith}},\ }\href {\doibase 10.1017/CBO9780511802850} {\emph {\bibinfo
  {title} {Bose-Einstein Condensation in Dilute Gases}}},\ \bibinfo {edition}
  {2nd}\ ed.\ (\bibinfo  {publisher} {Cambridge University Press},\ \bibinfo
  {address} {Cambridge},\ \bibinfo {year} {2008})\ Chap.~\bibinfo {chapter}
  {12}\BibitemShut {NoStop}%
\bibitem [{\citenamefont {Dalfovo}\ \emph {et~al.}(1999)\citenamefont
  {Dalfovo}, \citenamefont {Giorgini}, \citenamefont {Pitaevskii},\ and\
  \citenamefont {Stringari}}]{DalfovoRev1999}%
  \BibitemOpen
  \bibfield  {author} {\bibinfo {author} {\bibfnamefont {F.}~\bibnamefont
  {Dalfovo}}, \bibinfo {author} {\bibfnamefont {S.}~\bibnamefont {Giorgini}},
  \bibinfo {author} {\bibfnamefont {L.~P.}\ \bibnamefont {Pitaevskii}}, \ and\
  \bibinfo {author} {\bibfnamefont {S.}~\bibnamefont {Stringari}},\ }\href
  {\doibase 10.1103/RevModPhys.71.463} {\bibfield  {journal} {\bibinfo
  {journal} {Rev. Mod. Phys.}\ }\textbf {\bibinfo {volume} {71}},\ \bibinfo
  {pages} {463} (\bibinfo {year} {1999})}\BibitemShut {NoStop}%
\bibitem [{\citenamefont {Jackson}\ \emph {et~al.}(1998)\citenamefont
  {Jackson}, \citenamefont {McCann},\ and\ \citenamefont
  {Adams}}]{Jackson1998}%
  \BibitemOpen
  \bibfield  {author} {\bibinfo {author} {\bibfnamefont {B.}~\bibnamefont
  {Jackson}}, \bibinfo {author} {\bibfnamefont {J.~F.}\ \bibnamefont {McCann}},
  \ and\ \bibinfo {author} {\bibfnamefont {C.~S.}\ \bibnamefont {Adams}},\
  }\href {http://stacks.iop.org/0953-4075/31/i=20/a=008} {\bibfield  {journal}
  {\bibinfo  {journal} {Journal of Physics B: Atomic, Molecular and Optical
  Physics}\ }\textbf {\bibinfo {volume} {31}},\ \bibinfo {pages} {4489}
  (\bibinfo {year} {1998})}\BibitemShut {NoStop}%
\end{thebibliography}%



\end{document}